\documentclass[
aip, 
jcp, 
reprint,
floatfix]{revtex4-1} %
\usepackage{graphicx}
\usepackage[colorlinks=true, linkcolor=green, citecolor=red]{hyperref}
\usepackage{xcolor}
\usepackage{amsmath}
\usepackage{graphicx}
\usepackage{xcolor}
\usepackage{setspace}
\usepackage{physics}
\usepackage{amsfonts}
\usepackage{float}
\usepackage{tikz}
\usepackage{dirtytalk}
\usetikzlibrary{shapes.geometric, positioning, arrows.meta}
\usepackage{adjustbox}  %
\usepackage[final]{pdfpages}
\usepackage{pgffor}

\renewcommand{\selectlanguage}[1]{}

\makeatletter
\def\fps@figure{t}
\AtBeginDocument{\let\LS@rot\@undefined}
\makeatother

\begin{document}

	\title{The Good, the Bad, and the Ugly of Atomistic Learning for ``Clusters-to-Bulk'' Generalization}
	\author{Miko{\l}aj J. Gawkowski}
	\affiliation{%
		Department of Physics and Astronomy, University College London, 7-19 Gordon St, London WC1H 0AH, UK
	}
	\affiliation{%
		Thomas Young Centre and London Centre for Nanotechnology, 
        9 Gordon St, London WC1H 0AH, UK
	}

	\author{Mingjia Li}
	\affiliation{%
		Department of Physics and Astronomy, University College London, 
        7-19 Gordon St, London WC1H 0AH, UK
	}
	\affiliation{%
		Thomas Young Centre and London Centre for Nanotechnology, 
        9 Gordon St, London WC1H 0AH, UK
	}

        \author{Benjamin X. Shi}
	\affiliation{%
		Initiative for Computational Catalysis,        
        Flatiron Institute, New York, New York 10010, USA
	}

	\author{Venkat Kapil}%
	\email{v.kapil@ucl.ac.uk}

	\affiliation{%
		Department of Physics and Astronomy, University College London, 7-19 Gordon St, London WC1H 0AH, UK
	}
	\affiliation{%
		Thomas Young Centre and London Centre for Nanotechnology, 
        9 Gordon St, London WC1H 0AH, UK
	}

    \begin{abstract}

Training machine learning interatomic potentials (MLIPs) on total energies of molecular clusters using differential or transfer learning is becoming a popular route to extend the accuracy of correlated wave-function theory to condensed phases. 
A key challenge, however, lies in validation, as reference observables in finite-temperature ensembles are not available at the reference level. 
Here, we construct synthetic reference data from pretrained MLIPs and evaluate the generalizability of cluster-trained models on ice-Ih, considering scenarios where both energies and forces and where only energies are available for training. 
We study the accuracy and data-efficiency of differential, single-fidelity transfer, and multi-fidelity transfer learning against ground-truth thermodynamic observables. 
We find that transferring accuracy from clusters to bulk requires regularization, which is best achieved through multi-fidelity transfer learning when training on both energies and forces. 
By contrast, training only on energies introduces artefacts: stable trajectories and low energy errors conceal large force errors, leading to inaccurate microscopic observables. 
More broadly, we show that accurate reproduction of microscopic structure correlates strongly with low force errors but only weakly with energy errors, whereas global properties such as energies and densities correlate with low energy errors. 
This highlights the need to incorporate forces during training or to apply careful validation before production. 
Our results highlight the promise and pitfalls of cluster-trained MLIPs for condensed phases and provide guidelines for developing -- and critically, validating -- robust and data-efficient MLIPs.

\end{abstract}
	
\maketitle
	
\newpage

\section{Introduction}

Machine learning interatomic potentials (MLIPs) have emerged as powerful tools to reduce the computational cost of first-principles-level simulations across chemistry~\cite{kovacs_mace-off_2025}, materials science~\cite{batatia_foundation_2024, merchant_scaling_2023, yang_mattersim_2024}, and biology~\cite{unke_biomolecular_2024}.
Traditionally, MLIPs are trained on total energies and forces derived from density functional theory (DFT), inheriting the accuracy of DFT at a significantly reduced computational cost.
While this strategy has been highly successful, DFT lacks a systematic path to quantitative improvement, limiting its accuracy for many classes of systems~\cite{feibelmanCOPt1112001,baoSelfInteractionErrorDensity2018b,shi_general_2022,bryentonDelocalizationErrorGreatest2023,shiGoingGoldstandardAttaining2024a}.
Correlated wave-function theory (cWFT), by contrast, provides a hierarchy of systematically improvable methods that can approach experimental accuracy~\cite{schimkaAccurateSurfaceAdsorption2010b,yangInitioDeterminationCrystalline2014,sauerInitioCalculationsMolecule2019b,ye_adsorption_2024,della_pia_how_2024, shi_accurate_2025}.
Training MLIPs directly on cWFT data is therefore highly desirable, but poses many challenges: the steep computational cost of cWFT calculations, the high cost and complexity of computing energy gradients, and the limited availability of periodic implementations, which often requires training MLIPs for bulk systems on molecular clusters~\cite{daru_coupled_2022, meszaros_short-range_2025, oneill_towards_2025}. 
We refer to this process as “cluster-to-bulk” generalization.

Several strategies have been proposed to reach cWFT-level accuracy in MLIPs, with liquid water and ice serving as prototypical systems.
Early “machine-learning” approaches were rooted in the many-body expansion, with the water molecule as the basic unit, leading to accurate water models such as MBpol~\cite{medders_representation_2015, reddyAccuracyMBpolManybody2016} and, more recently, q-AQUA(-pol)~\cite{quInterfacingQAQUAPolarizable2023}.
These models are based on permutationally invariant polynomials~\cite{xie_permutationally_2010}, which learn up to three- or four-body contributions, and a semi-empirical polarizable force field which approximates higher-order terms~\cite{medders_representation_2015}.
In contrast, molecule-agnostic approaches have directly applied MLIPs.
In this context, Lan et al.~\cite{lan_simulating_2021} carried out a \textit{tour de force} study, based on the Behler–Parrinello framework~\cite{behler_constructing_2015}, directly training on resolution-of-identity MP2~\cite{del_ben_electron_2013} total energies and forces, which required more than 20,000 periodic configurations.

Considering the high data requirement of training on cWFT data with standard MLIPs, many strategies have been proposed.
The first class of strategies eases the learning task by using DFT as an intermediate level of theory.
For example, \citet{chen_data-efficient_2023} adopted a transfer learning strategy~\cite{Thrun1995}: the MLIP was first pretrained on DFT-level energies and forces, then fine-tuned on a much smaller dataset of cWFT-level energies using the pretrained weights for initialization.
This approach proved data-efficient for CCSD~\cite{purvis_full_1982}, CCSD(T)~\cite{raghavachari_fifth-order_1989}, and auxiliary-field quantum Monte Carlo~\cite{zhang_quantum_2003}, requiring fewer than 200 total-energy evaluations for periodic configurations with up to 100 atoms. 
However, validation was restricted to a single thermodynamic state point in the $NVT$ ensemble.
In a complementary effort, \citet{daru_coupled_2022} introduced a differential learning framework~\cite{ramakrishnan_big_2015}, which was further supplemented by a fixed point-charge Coulomb baseline and short-range repulsion terms.
Here, one model was trained on total energies and forces at a local MP2 level, while a second captured the energy differences between local MP2~\cite{pinski_sparse_2015} and local CCSD(T)~\cite{riplinger_sparse_2016}.
Crucially, training relied on nearly 20,000 small water clusters carved from periodic MLIP simulations, substantially reducing the computational and memory demands of generating cWFT data.
Validation was again limited to liquid water in the $NVT$ ensemble at a single state point.

A second class of approaches focuses on scalable MLIP architectures that intrinsically require less training data.
Iterative schemes that construct high-body-order features from tensor products of low-body-order terms~\cite{nigam_recursive_2020}, most prominently within the atomic cluster expansion (ACE) framework~\cite{drautz_atomic_2019}, enable systematically improvable and data-efficient models.
Graph neural network architectures such as NequIP~\cite{nequip}, MACE~\cite{batatia_design_2022}, and GRACE~\cite{bochkarev_graph_2024} extend the scalability of ACE features across chemical space through tensor reduction~\cite{darby_tensor-reduced_2023}.
Building on this, \citet{kaur_data-efficient_2024} demonstrated that transfer learning, via fine-tuning of the so-called MACE-MP-0 model~\cite{batatia_foundation_2024} pretrained on the diverse “MPTrj” dataset~\cite{https://doi.org/10.6084/m9.figshare.23713842}, further reduces data requirements compared to earlier architectures and direct training with the MACE architecture.
Using fewer than 50 periodic configurations with up to 100 molecules, they achieved sub-1\% density errors across four ice phases at finite thermodynamic conditions, enabling RPA-quality simulations in the NPT ensemble.
More recently, \citet{meszaros_short-range_2025} combined ACE models with differential learning to train MLIPs for liquid water on water clusters carved from the bulk liquid. 
The ACE descriptors reduced the required training data from the 20,000 clusters reported in their earlier work~\cite{daru_coupled_2022} to just 1,000 clusters, while working in the $NVT$ ensemble. 
Building upon this, \citet{oneill_towards_2025} incorporated clusters of increasing radii from liquid water structures from the $NPT$ ensemble and fitting a differential learning model to the difference between local CCSD(T) and DFT energies, achieving good agreement with experiments.
Independently, more advanced transfer learning strategies, notably multi-fidelity fine-tuning amongst others~\cite{bocus_operando_2025, novelli_fast_2025, radova_fine-tuning_2025, mazitov_pet-mad_2025}, have been studied and demonstrated to outperform differential learning, separately for both molecular~\cite{messerly_multi-fidelity_2025} and solid-state~\cite{cui_multi-fidelity_2025} systems; however, this approach remains to be tested on the cluster-to-bulk generalization of MLIPs.

These studies suggest that an affordable route to cWFT-level MLIPs combines scalable architectures with training on clusters, while leveraging differential or transfer learning in conjunction with an intermediate level of theory. 
However, a challenge associated with these approaches is their validation. 
Unlike DFT-based work, the ground-truth observables in finite thermodynamic ensembles -- against which the models should ultimately be benchmarked -- are prohibitive. 
As a result, validation relies either on reproducing the energies of clusters extracted from the bulk trajectory~\cite{daru_coupled_2022}, local convergence of observables with respect to the dataset, or on comparison with experiments~\cite{oneill_towards_2025}. 
All these routes have certain limitations.
The first assumes that a model that yields stable trajectories and low energy errors on clusters also yields low errors on forces and other microscopic observables.
Similarly, the criterion based on the local convergence of observables with respect to the dataset assumes that the MLIPs trained on clusters can be systematically improved to the ground truth. 
And finally, a comparison with experiments doesn't allow for disentangling error cancellation between the MLIP and the cWFT reference.  
A second important question is understanding how sensitive the cluster-to-bulk generalization of MLIPs is to the choice of training strategies, whether energies alone or both energies and forces are used, the volume of training data, and the size of clusters employed and their interplay.
In this work, we assess the ability of MLIPs trained on molecular clusters to generalize to bulk phases, using ice-Ih as a prototypical molecular solid. 
We evaluate four training strategies: (i) transfer learning via fine-tuning a pretrained model, (ii) differential learning with respect to a pretrained model, (iii) transfer learning with multi-fidelity fine-tuning, and (iv) direct learning on cluster data, which serves as a baseline. 
We benchmark these strategies across two scenarios: training on energies and forces versus training only on energies, assessing how well both global and microscopic observables are recovered in the $NVT$ and $NPT$ ensembles. 
Overall, we find that incorporating forces into training is highly beneficial, as it improves stability and enables simultaneous accuracy in both global properties (energies and density) and microscopic structure. 
Notably, multi-fidelity transfer learning performs best in this setting, reaching sub-meV/atom energy errors, densities within 1\% of the reference, and excellent agreement with microscopic structure using as few as 100 clusters. 
When trained only on energies, however, models reproduce global properties such as energy and density reasonably well, but consistently fail to capture microscopic structure accurately. 
In this case, differential learning performs best among the strategies considered. 
Our findings critically evaluate the cluster-to-bulk generalization of MLIPs and provide practical guidelines for developing and validating data-efficient MLIPs that can reliably generalize to condensed-phase systems.

The paper is organized as follows:
Section~\ref{s:methods} introduces the benchmark system, training strategies, datasets, and test metrics. 
Section~\ref{s:results} examines model performance by analyzing how well global and microscopic observables are recovered as a function of training data volume in the $NVT$ and $NPT$ ensembles. 
Section~\ref{s:discussion} summarizes the strengths and limitations of the different strategies in the context of the cluster-to-bulk generalization, while also considering extrapolation behavior and systematic improvement with increasing cluster size, and outlines directions for future work.  

\begin{figure*}[t]
    \centering
    \includegraphics[width=0.90\linewidth]{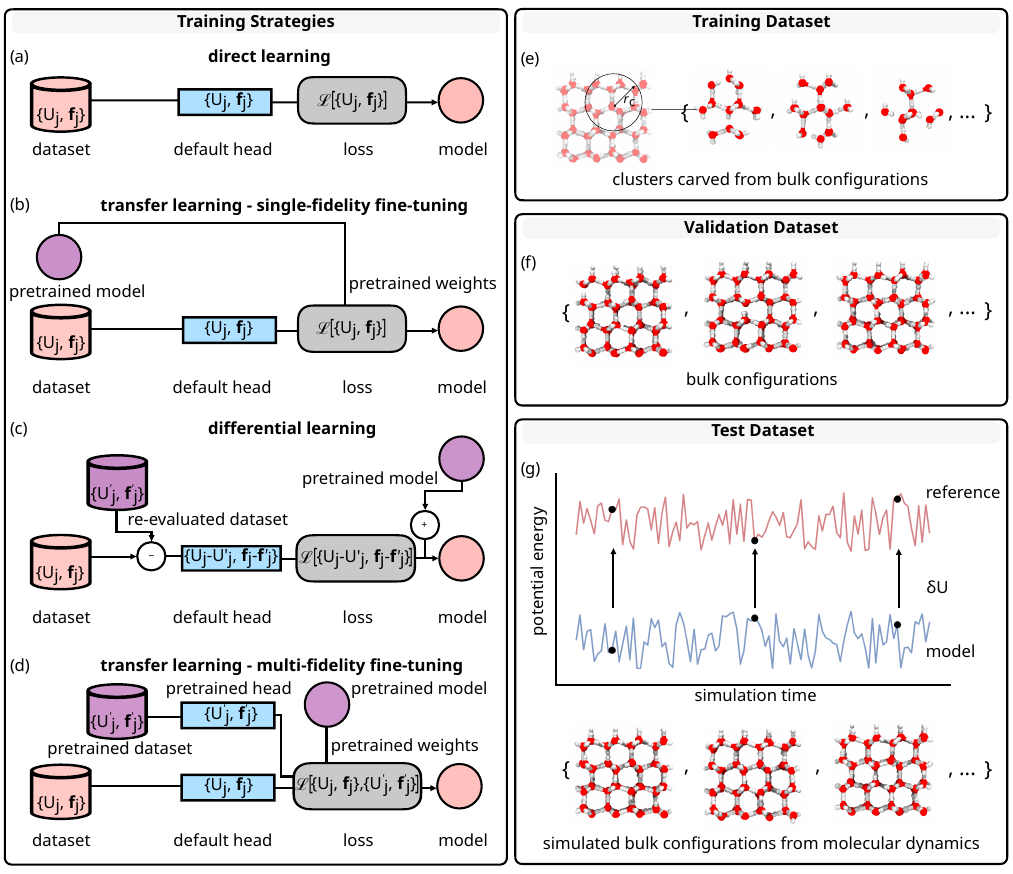}
    \caption{Schematic overview of the training strategies and datasets used in this work. Panels (a–d) illustrate the training strategies: direct learning, single-fidelity fine-tuning, differential learning, and multi-fidelity transfer learning, respectively. Panels (e-g) respectively show the datasets: clusters carved from bulk ice-Ih for training, bulk configurations for validation, and bulk configurations generated from molecular dynamics by the models for testing. In panel (c), the pretrained model serves as the baseline. 
}
    \label{fig:scheme}
\end{figure*}

\section{Methods}
\label{s:methods}

\subsection{System}

To assess the generalizability of MLIPs for bulk phases when trained on molecular clusters, we use ice-Ih as a benchmark system.
This system has proven to be a suitable testbed for studying the data efficiency of transfer learning~\cite{kaur_data-efficient_2024} via fine-tuning of pretrained MLIPs.
The guidelines derived from benchmarks for ice-Ih were shown to apply to ice phases and the X23 dataset of rigid molecular crystals~\cite{pia_accurate_2025}.
Thus, we expect the insights gained from this study to apply to other molecular systems. 

\subsection{Training strategies for MLIPs}

We discuss four training strategies that combine high- and low-fidelity data or levels of theory differently. 
A qualitative overview of these strategies is shown in Fig.~\ref{fig:scheme}(a)-(d).
Here, high-fidelity level refers to the target level of theory, and the corresponding high-fidelity data consists of labelled energies and forces at this level. 
Analogously, the low-fidelity level refers to an intermediate theory -- ideally much less expensive yet semi-quantitatively accurate-- that can be leveraged to improve data efficiency relative to training solely on high-fidelity data. \\

\textbf{Direct learning}
corresponds to standard regression, where the MLIP is trained from randomly initialized weights solely on high-fidelity data as depicted in Fig.~\ref{fig:scheme}(a). 
This is achieved by minimizing a loss function that measures the discrepancy between the reference values and the model predictions.
For the set of $M$ training configurations $\{\mathbf{q}_{j}\}_{j=1}^{M}$  and the set of the corresponding high-fidelity total energies and forces, $\{U_{j}, \mathbf{f}_{j}\}_{j=1}^{M}$, the loss is defined as 
\begin{align}
    \mathcal{L}^{\text{direct}} & = \mathcal{L}\left[\{U_{j}, \mathbf{f}_{j}\}_{j=1}^{M}\right]   = \ \frac{1}{M} \sum_{j=1}^{M} \Bigg[  \lambda_{U} \left(U_j{} -\hat{U}\left(\mathbf{q}_{j}\right)\right)^2 \nonumber\\ &  +  \lambda_{\mathbf{f}} \left|\left|\mathbf{f}_{j} - \hat{\mathbf{f}}(\mathbf{q}_j)\right|\right|_{2}^2 \Bigg]
\end{align}
with $\lambda_{U}$ and $\lambda_{\mathbf{f}}$ are referred to as the energy and the force weights respectively, $\hat{U}$ is the potential energy predicted by the MLIP (the explicit dependence on the unit cell, the atomic identities, and the learnable weights are not shown for brevity), $\hat{\mathbf{f}}$ is the force predicted by the model (the MLIPs considered in this work are conservative with $\hat{\mathbf{f}} = -\nabla_{\mathbf{q}} \hat{U}$), and $\left|\left|\cdot\right|\right|_2$ corresponds to the $l^{2}$ norm.
In this work, the high-fidelity training data comprises total energies and forces estimated on water cluster configurations carved from bulk structures. \\

\textbf{Transfer learning with single-fidelity fine-tuning} involves two steps: pretraining of the MLIP on low-fidelity training data, followed by fine-tuning of the same model on high-fidelity training data.
For a low-fidelity dataset consisting of a set of $M'$ configurations $\{\mathbf{q}'_{j}\}_{j=1}^{M'}$ and their total energies and forces $\{U'_{j}, \mathbf{f}'_{j}\}_{j=1,}^{M'}$, the pretraining step involves direct training of an MLIP, by minimizing the low-fidelity loss
\[\mathcal{L}^{\text{direct}}_{\text{low-fidelity }} = \mathcal{L}\left[\{U'_{j}, \mathbf{f}'_{j}\}_{j=1}^{M'}\right].\]
The final weights of this pretrained model will be referred to as pretrained weights.
Low-fidelity data can comprise larger and more complex configurations, including the target bulk phase.

In the single-fidelity fine-tuning step, as depicted in Fig.~\ref{fig:scheme}(b), the MLIP is initialized to the pretrained weights and optimized on a new loss defined for the high-fidelity training data comprising molecular clusters
\[\mathcal{L}^{\text{direct}}_{\text{high-fidelity }} = \mathcal{L}\left[\{U_{j}, \mathbf{f}_{j}\}_{j=1}^{M}\right].\] 

\textbf{Differential learning} learns the difference between high- and low-fidelity levels for a given dataset \cite{ramakrishnan_big_2015}, as depicted in Fig.~\ref{fig:scheme}(c).
It requires a low-fidelity model, which need not share the same architecture as the MLIP being trained, or even be an MLIP.
In the first step, the $M$ high-fidelity configurations $\{\mathbf{q}_{j}\}_{j=1}^{M}$ are evaluated using the low-fidelity model to obtain the corresponding low-fidelity energies and forces, $\{U'_{j}, \mathbf{f}'_{j}\}_{j=1}^{M}$.
A new set of target values is constructed
\[\{\delta U_{j}, \delta \mathbf{f}_{j}\}_{j=1}^{M} \equiv \{U_{j} - U'_{j}, \mathbf{f}_{j} - \mathbf{f}'_{j}\}_{j=1}^{M}
\]
representing the difference between the high- and low-fidelity predictions.
The MLIP is trained, starting from randomly initialized model weights, to minimize the loss
\[
\mathcal{L}^{\text{differential}} = \mathcal{L}\left[\{\delta U_{j}, \delta \mathbf{f}_{j}\}_{j=1}^{M}\right]
\]
The final model prediction is obtained by summing the low-fidelity model and the MLIP: $\hat{U} = U' + \delta\hat{U}$ with $\delta\hat{U}$ representing the differential model. \\

\textbf{Transfer learning with multi-fidelity fine-tuning} utilizes low-fidelity data during both the pretraining and fine-tuning stage, as depicted schematically in Fig.~\ref{fig:scheme}(d).
In the first step, the MLIP is first pretrained on low-fidelity training data.
To facilitate the explanation of the rest of the approach, we introduce terminology for different components of the MLIPs. 
The output block of the pretrained model, a small multilayer perceptron that predicts low-fidelity energies and forces $\hat{U}', \hat{\mathbf{f}}'$, will be referred to as the pretraining (pt) head. 
The remainder of the network will be referred to as the shared layers.
During the fine-tuning stage, the model is augmented with an additional output block tasked with predicting high-fidelity energies and forces $\hat{U}, \hat{\mathbf{f}}$, referred to as the default head.
With two output heads, the model can be trained simultaneously on low- and high-fidelity data by minimizing a weighted sum of their corresponding direct losses
\begin{equation}
\mathcal{L}^{\text{multi-fidelity}} = \lambda_{\text{pt}} ~\mathcal{L}^{\text{direct}}_{\text{low-fidelity}} + \lambda_{\text{default}} ~\mathcal{L}^{\text{direct}}_{\text{high-fidelity}}
\end{equation}
where $\lambda_{\text{pt}}$ and $\lambda_{\text{default}}$ control the relative importance of the two loss contributions.
The model weights are initialized using the pretrained values for the shared layers and both the pretraining and default heads.
The multi-fidelity approach offers flexibility in dataset selection -- for example, enabling pretraining on a diverse low-fidelity dataset and fine-tuning on a more specialized low-fidelity dataset that closely resembles the target system.
In the context of the cluster-to-bulk generalization of MLIPs, a desirable option that leverages the computational advantages of both fidelity levels is to use low-fidelity bulk configurations with total energies and forces alongside high-fidelity total energies for clusters.

\subsection{Datasets and test metrics}

\textbf{Levels of theory}:
Although the ultimate goal of this work is to evaluate training strategies for cWFT methods, we employ pretrained MLIPs as inexpensive proxies. 
This step is essential, since extensive benchmarking in finite thermodynamic ensembles against known ground truth values would be computationally prohibitive with cWFT methods, making it otherwise impossible to assess the accuracy of the cluster-to-bulk training paradigms.

We select the MACE-MP-0 model, trained at the generalized gradient approximation density functional PBE, as the low-fidelity level, and the MACE-OFF-23 model, trained at the van der Waals (vdW)-inclusive hybrid-functional level $\omega$B97M-D3(BJ), as the high-fidelity level. 
We recognize that pretrained MLIP proxies may introduce artifacts, owing to their locality and the use of training data restricted to clusters or small unit cells. 
To address this, we validate our conclusions with explicit DFT calculations, which confirm that the trends observed with proxy MLIPs also hold for true electronic structure calculations, as shown in Fig. S3 of the Supporting Information (SI). 
\\

\textbf{Training and validation datasets:} The high-fidelity training sets consist of total energies and forces for molecular clusters. 
As shown in Fig.~\ref{fig:scheme}(e), these clusters are carved from bulk configurations.
To systematically explore data efficiency, we construct subsets of increasing size, containing 50, 100, 200, 400, 800, and up to 1000 clusters for each cluster size defined by the cluster radius $r_{\text{c}} \in \{4.0, 4.5, 6.0\}$ \AA{}. 
Most of the results presented in this work are based on $4.5$ \AA{} sized clusters containing 15 water molecules on average.

The low-fidelity training data comprises two types.
The first type corresponds to a subset of configurations sampled from the MPTrj -- the dataset used to train the MACE-MP-0 model.
The second type corresponds to a set of ice-Ih configurations extracted from molecular dynamics~\cite{kaur_data-efficient_2024}.
For differential learning -- which does not require low-fidelity training data per se, but rather a low-fidelity model -- we use the MACE-MP-0 model as the reference baseline. 
The validation dataset consists of 25 periodic ice-Ih configurations evaluated at the high-fidelity level of theory (see Fig.~\ref{fig:scheme}(f)).
For the isolated atomic energies, we use the MACE-OFF-23 isolated atomic energy values of $U_{H}$ = -13.571964772646918 eV, $U_{O}$ = -2043.933693071156 eV where $U_{H}$ is the energy of hydrogen and $U_{O}$ of oxygen and in the case of differential learning, where one trains on differences in energies and forces, we use $\delta U_{H}$ = -12.412976535265937 eV, $\delta U_{O}$ = -2041.8522316415601 eV, where each energy is the difference between MACE-OFF-23 and MACE-MP-0 references.
\\

\textbf{Test ensembles and metrics}:
Building on our earlier observation that validation errors do not reflect out-of-domain performances in molecular dynamics ensembles~\cite{kaur_data-efficient_2024}, we evaluate each model’s ability to generalize by performing molecular dynamics simulations in the $NVT$ and $NPT$ thermodynamic ensembles.
In the $NVT$ ensemble, we report the energy errors on configurations sampled from trajectories generated by the MLIPs, as depicted in Fig.~\ref{fig:scheme}(g). 
While these configurations are not drawn from the high-fidelity distribution, the resulting energy error can be interpreted -- via thermodynamic perturbation theory -- as a measure of the MLIP's ability to represent the true ensemble~\cite{tuckerman_statistical_2010}, provided it is interpreted with care.
In the $NPT$ ensemble, in addition to energy errors, we also report the predicted density. 
Density errors are computed relative to the reference density obtained from molecular dynamics simulations performed using the high-fidelity MACE-OFF-23 model under identical thermodynamic conditions.
In Fig.~S1 of the SI, we report the density, the O--O pair correlation function, and the O--H bond distribution at both high- and low-fidelity levels.

\subsection{Computational details}

\textbf{Machine learning:} The MACE architecture~\cite{batatia_mace_2022} is used to represent the MLIPs, and all numerical experiments are performed using version 0.3.9.
Unless stated otherwise, the MACE models used in this work correspond to the ``medium" architecture from Ref.~\citenum{batatia_foundation_2024} and are trained in 64-bit floating-point precision.
Training is performed using the AMSGrad variant of the Adam optimizer with default parameters: $\beta_1 = 0.9$, $\beta_2 = 0.999$, and $\epsilon = 10^{-8}$, for 150 epochs.
We use a learning rate of 0.001, an exponential moving average scheduler with a decay factor of 0.995, and apply gradient clipping with a maximum norm of 100.
After 150 epochs, and 500 for differential learning, the energy loss weight was increased by 20\%, and training continued for an additional 150 and 500 epochs, respectively. 
The larger number of epochs for differential learning was used to account for its slower learning dynamics. 
The final model is selected from the checkpoints (across both training stages) that minimize the validation loss. \\

\textbf{Molecular dynamics:} All molecular dynamics simulations are performed using the i-PI code ~\cite{ipi}, using an ASE client~\cite{ase}, for calculating total energy and forces for the MLIPs. 
All simulations are thermalized using a Langevin thermostat with a time constant of 100\,fs; for $NPT$ simulations, we utilize the fully flexible Martyna-Tuckerman-Tobias-Klein barostat with a relaxation time of 1 ps, and both the system and the barostat ``piston" are coupled with a Langevin thermostat with a time constant of 100\,fs.
We use a timestep of 0.5 fs, and sample positions, potential energies, and densities every 200 timesteps for 50\,ps.
The first 10 ps of the simulations are discarded as equilibration.

\begin{figure*}[t]
    \centering
    \includegraphics[width=0.90\linewidth]{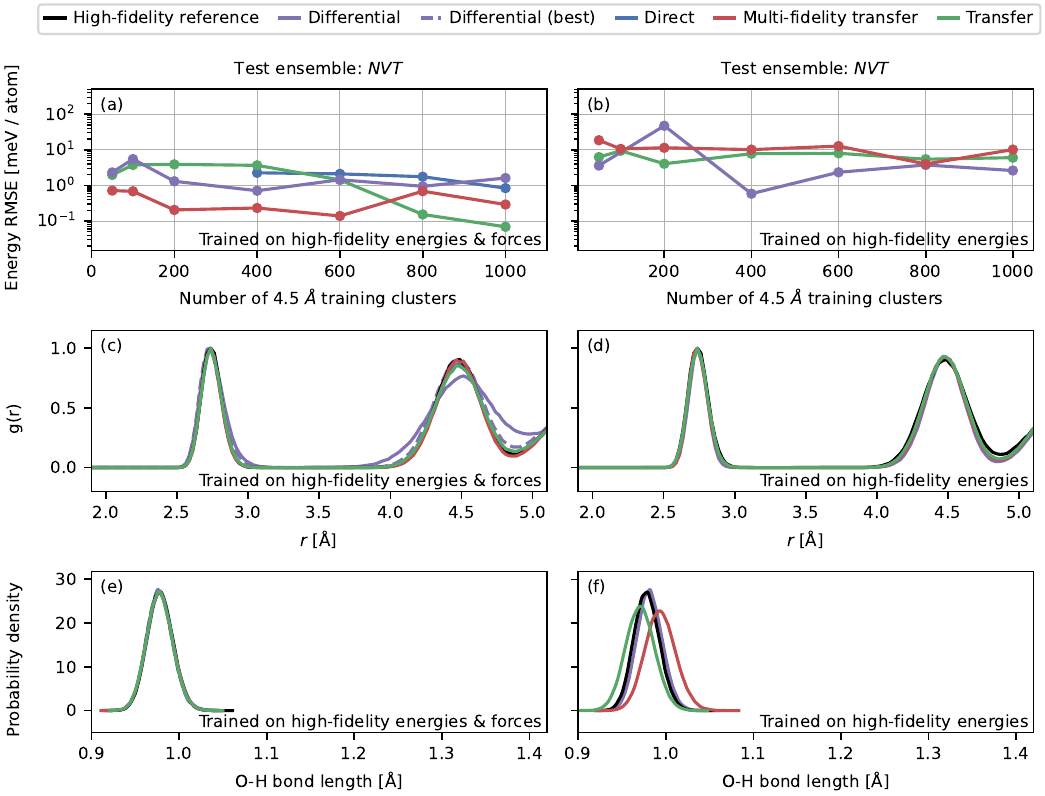}
    \caption{Panels (a,c,e) show $NVT$ energy RMSEs, O-O pair correlation functions, and O-H bond distributions for models trained on high-fidelity energies and forces. Panels (b,d,f) show the same quantities for models trained on high-fidelity energies only. `High-fidelity reference' corresponds to the MACE-OFF-23 model. `Differential (best)' refers to the delta learning model that achieved the lowest energy RMSE and was trained on 400 clusters. `Transfer' is the single-fidelity fine-tuning method.}
    \label{fig:nvt-testing}
\end{figure*}

\section{Results}
\label{s:results}
Before testing the performance of the models in the thermodynamic ensembles, we report total energy and force validation RMSEs across all strategies as a function of the training set size (number of 4.5\,\AA{} clusters) in Fig.~S2 of the SI. 
The left panels correspond to training with high-fidelity forces included, and the right panels to training without forces.
To arrive at these results, we perform for each strategy, a hyperparameter sweep over energy and force weights, with $\lambda_{U}, \lambda_{\mathbf{f}} \in \{0.001, 0.01, 0.1, 1, 10, 100, 1000\}$, and report the RMSEs corresponding to the models that minimize the validation error.
Hyperparameter optimization yields more than an order-of-magnitude improvement in energy and force RMSEs, with the optimal weights differing across training strategies. 
Table~S1 of the SI reports the optimal weights used across the different strategies.

\subsection{Test errors in the $NVT$ ensemble}

The accuracy and data efficiency of the training strategies in the $NVT$ ensemble are summarized in Fig.~\ref{fig:nvt-testing}. 
Results are shown separately for two scenarios: training on high-fidelity energies and forces, and training on high-fidelity energies (i.e., with forces omitted). 
As a first step, we report an error metric on a global observable -- the average RMSE associated with potential energy. 
This metric is fundamental: in practice, the total energies of bulk systems or clusters carved from them at cWFT levels are the only reference data for validating a model~\cite{daru_coupled_2022, chen_data-efficient_2023, oneill_towards_2025}. 
We further examine microscopic observables with respect to the ground truth thermodynamic averages, namely the first two peaks of the O--O pair correlation function and the distribution of O--H bond lengths. 
This assessment is important, as average structural observables are often inaccessible at high-fidelity levels; it is typically assumed that low energy RMSEs are a sufficient condition for the identification of an accurate model. \\

\textbf{Global observable:} In panels (a) and (b) of Fig.~\ref{fig:nvt-testing}, we report the energy RMSE across $NVT$ trajectories generated by the MLIPs, as a function of the number of clusters in the training set. 
We first consider the scenario when high-fidelity forces are included in training. 
As shown in Fig.~\ref{fig:nvt-testing}(a), all methods achieve sub- or near-meV/atom energy RMSE for the largest training set volume.
Direct learning yields unstable trajectories and high RMSEs for small training sets, but improves systematically and achieves a sub-meV/atom RMSE with 1000 configurations. 
Differential learning yields unstable trajectories for 100 clusters, but with more training data, shows improvement in simulation stability, although the energy RMSEs remain saturated around 1\,meV/atom.
Transfer learning approaches lead to stable simulations across the full data range, with single-fidelity fine-tuning leading to larger errors for small dataset sizes and eventually reaching the lowest energy error for 1000 clusters.
Multi-fidelity learning yields near-constant sub-meV/atom errors, even with just 50 clusters. 

We next consider the scenario when high-fidelity forces are not included in the training. 
Results from the direct learning are omitted in Fig.~\ref{fig:nvt-testing}(b), as the trajectories were unstable. 
This occurs despite these models reporting systematically improving learning curves on the validation set, and achieving a sub-meV/atom energy RMSE for 1000 clusters (see Fig.~S2 of the SI).
Differential learning is unstable when training with 100 clusters, but improves with more data, saturating at $\sim$2\,meV/atom with 1000 configurations. 
Standard transfer learning maintains stable performance across all training set sizes, reaching $\sim$5\,meV/atom error. 
Multi-fidelity transfer learning, by contrast, yields stable simulations but consistently higher errors of $\sim$10\,meV/atom. \\

\textbf{Microscopic observables:} 
We now examine if low-energy RMSEs correlate with a good description of the microscopic structure.
We begin with the scenario in which high-fidelity forces are included in training. 
Figure~\ref{fig:nvt-testing}(c) shows the oxygen–oxygen pair correlation functions up to the second peak for models trained on 1000 clusters. 
Direct learning and both transfer learning approaches (single-fidelity and multi-fidelity fine-tuning) reproduce the reference distribution within statistical error. 
By contrast, differential learning yields broadened first and second peaks. 
It is important to note that the differential learning model trained with 1000 configurations is not the one that achieves the lowest energy RMSE. 
For this reason, we also examine the best-performing differential learning model trained on 400 configurations.
This model shows an improved agreement with the reference but still exhibits a mild systematic broadening. 
Turning to the O--H bond length distributions in Fig.~\ref{fig:nvt-testing}(e), all models—including both differential learning models—accurately reproduce the reference distribution. 

We next consider the scenario in which forces are not included in training. 
As shown in Fig.~\ref{fig:nvt-testing}(d), all strategies (with the exception of direct learning, which yields unstable trajectories) reproduce the first peak of the O–O pair correlation function with quantitative accuracy. 
The differential learning model trained only on energies outperforms its counterpart, which is trained on both energies and forces. 
This suggests that, for differential learning, including forces in the training set leads to overfitting on clusters, hindering generalization to the bulk phase.
For the second peak, however, all methods exhibit a mild over-structuring. 
The O–H bond length distributions in Fig.~\ref{fig:nvt-testing}(f) reveal further issues. 
All approaches perform poorly for this O--H mean values and fluctuations. 
Here, both transfer learning approaches predict unphysical bond lengths, with single-fidelity fine-tuning yielding shorter and multi-fidelity fine-tuning yielding longer bond lengths. 
In contrast, differential learning shows the best performance but systematically overestimates the bond length by ~0.003\,\AA{}. 
As an absolute, this deviation is small, but it is in the same ballpark as the changes in the O--H bond length when nuclear quantum effects are incorporated~\cite{stolte_nuclear_2024}. 
These differences can modulate O--H autoionization~\cite{ceriotti_nuclear_2013, dasgupta_nuclear_2025} and lead to changes in the vibrational spectra by over 100 cm$^{-1}$~\cite{kapil_first-principles_2023}.
A possible explanation for the poor description of O--H modes is that when training on forces -- which fluctuate strongly for high-frequency modes as opposed to the energy -- the loss effectively assigns greater weight to these modes~\cite{bigi_dark_2025}. 
When training only on energies, this weighting is absent, which explains the systematically worse performance for high-frequency O--H compared to low-frequency O--O distributions.

These results may seem puzzling given the low energy RMSEs and stable trajectories; however, they are readily explained by the force RMSEs (calculated across the trajectories) shown in Fig.~S4(a) and S4(b). 
Training only on high-fidelity energies (a global quantity) can lead to low-energy RMSEs but high-force RMSEs, which correlates with the poor description of microscopic properties.
These findings highlight several caveats for the cluster-to-bulk generalization of MLIPs, which are further tested in the $NPT$ ensemble. 
(1) Stable trajectories and low energy errors do not imply accurate forces,  
(2) Agreement on global properties, such as total energy, does not guarantee correct microscopic structure.

\begin{figure*}[]
    \centering
    \includegraphics[width=\linewidth]{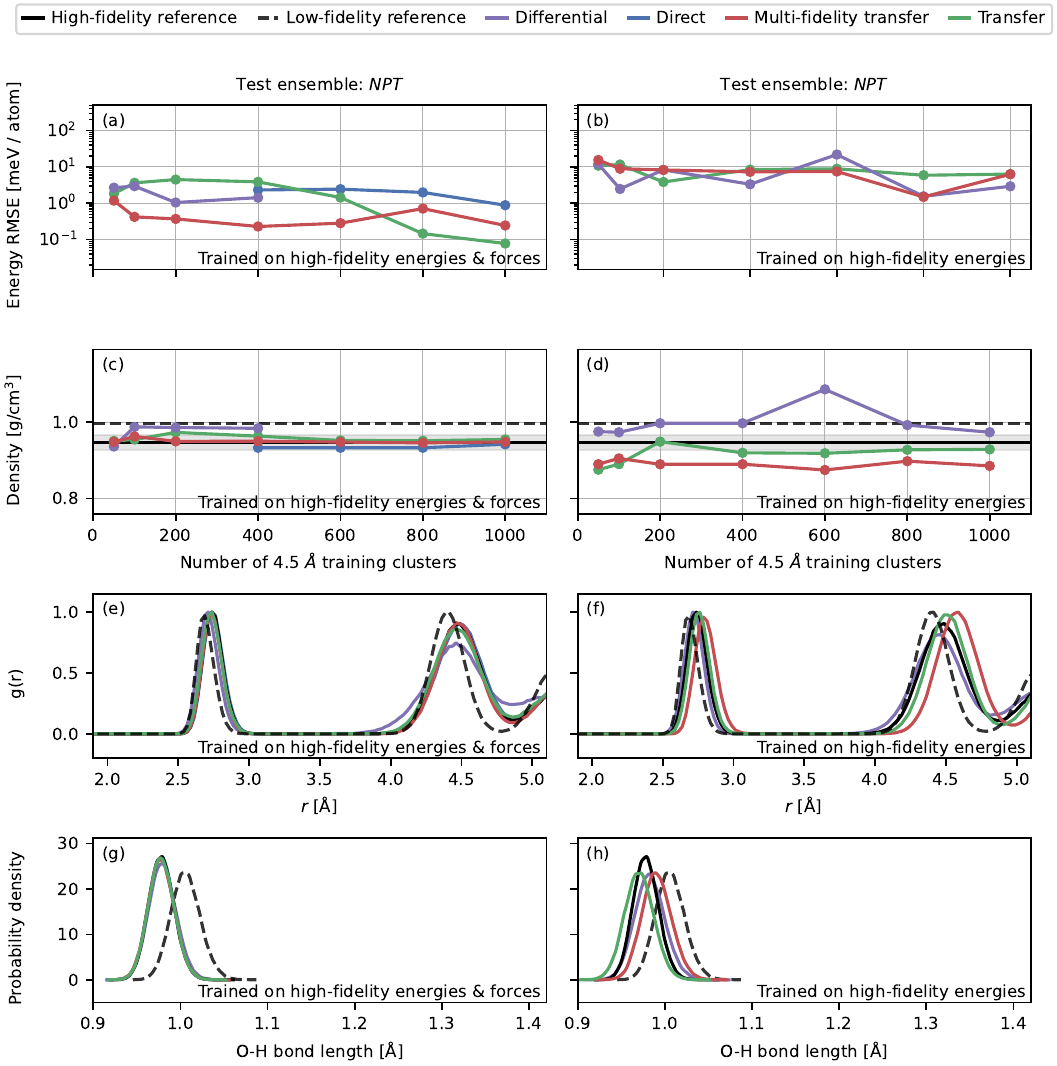}
    \caption{Panels (a,c,e,g) show $NPT$ energy RMSEs, average densities, O-O pair correlation functions, and O–H bond length distributions for models trained on high-fidelity energies and forces. Panels (b,d,f,h) show the same quantities for models trained on high-fidelity energies only. `Low-fidelity reference' is the MACE-MP-0 reference.}
    \label{fig:npt-testing}
\end{figure*}

\subsection{Test errors in the $NPT$ ensemble}

We next evaluate model performance in the $NPT$ ensemble. 
This represents a more stringent test ensemble, as energy and force errors directly affect the instantaneous pressure, which in turn can impact the stability of the simulation as well as global and microscopic observables. 
Moreover, it allows us to assess another important thermodynamic quantity, the density. \\

\textbf{Global observables:} 
In Fig.~\ref{fig:npt-testing}(a) and Fig.~\ref{fig:npt-testing}(b), we report the energy RMSEs across $NPT$ trajectories generated by the MLIPs, as a function of the number of clusters used for training. 
We first consider the scenario in which training includes both energies and forces. 
As shown in Fig.~\ref{fig:npt-testing}(a), direct models are unstable up to 400 clusters but improve systematically with larger training sets, reaching sub-meV/atom energy RMSEs. 
Differential learning models do not exhibit instability in the $NPT$ ensemble but clearly show signs of overfitting: the model trained on 200 clusters achieves the lowest RMSE, while adding more data leads to unstable simulations. 
Both transfer learning approaches achieve the lowest-energy RMSEs, with multi-fidelity fine-tuning performing better in the low-data regime, and single-fidelity fine-tuning outperforming at larger training set sizes.
We also consider the density as another global observable. 
In this case, we also contextualize the performance of the models with respect to the low-fidelity reference density.
As shown in Fig.~\ref{fig:npt-testing}(c), direct learning and both transfer learning strategies predict densities to within 2\% of the reference. 
Differential learning, on the other hand, leads to a systematic overestimation of the density up to 400 clusters (this value is close to the low-fidelity reference), and unstable simulations occur beyond this point. 

We next consider the scenario when high-fidelity forces are not included in the training.  
As shown in Fig.~\ref{fig:npt-testing}(b), the overall energy RMSEs are higher compared to the scenario where forces are incorporated in the training. 
This secondary effect is induced by errors in sampling, as understood by the high force RMSEs in Figs. S4(c) and S4(d). 
Direct learning leads to unstable simulations across the full range of training data.
Both types of transfer learning saturate in the range of 5-10 meV/atom errors.
Differential learning leads to energy RMSEs in the 2-11 meV/atom range.
We next analyze the density in Fig.~\ref{fig:npt-testing}(d), while comparing with the low-fidelity ground truth density estimate to understand if the errors arise from an undercorrection or an overcorrection. 
The description of the density is overall worse compared to the scenario when forces are included in the training. 
Differential learning leads to an overestimation of the density, to nearly the same extent as when training on both energies and forces, and close to the low-fidelity estimate.
This suggests that differential learning undercorrects. 
Single-fidelity fine-tuning leads to sub-2\% agreement with the reference, while multi-fidelity fine-tuning leads to an underestimation of the density, which suggests both these approaches (over)correct for the density difference between the two fidelity levels. \\

\textbf{Microscopic observables:} 
We first consider the scenario in which forces are included in the training.
Fig.~\ref{fig:npt-testing}(e) shows the oxygen–oxygen radial distribution functions along with the low- and high-fidelity references.
For the first peak of the distribution, both direct and transfer learning approaches yield a quantitative agreement with the reference. 
On the other hand, the best differential learning model (trained on 200 clusters) leads to a small underestimation of the nearest O--O distance. 
This deviation is in the same direction as the low-fidelity distribution, again implying that differential learning undercorrects.
For the second peak, direct learning and multi-fidelity fine-tuning yield a quantitative agreement, while single-fidelity fine-tuning and differential learning yield a distribution skewed towards the low-fidelity estimate. 
Fig.~\ref{fig:npt-testing}(g) shows the O--H bond distribution with all methods yielding quantitative agreement with the expectation of differential learning, which exhibits mild overstructuring. 
These results show that the accuracies achieved by the different approaches when training with forces in the $NVT$ ensemble carry over to the $NPT$ ensemble, and the artifacts are more pronounced.

We next consider the scenario in which high-fidelity forces are excluded from the training, starting the O--O distribution. 
From the $NVT$ analysis, we already anticipate a poor performance, but we check further if the low-fidelity level biases the microscopic structure.
As shown in ~\ref{fig:npt-testing}(f), the distribution of both the first and second peaks deviates from the reference for all the approaches. 
Single-fidelity fine-tuning skews the distributions to longer O--O distances, implying an overcorrection.
The multi-fidelity fine-tuning method skews the distributions towards longer distances. 
Differential learning shows the smallest errors, but systematically skews the distributions towards the low-fidelity reference. 
For the O--H distances, shown in Fig.~\ref{fig:npt-testing}(h), we note that differential learning and the multi-fidelity fine-tuning models are skewed towards the  
low-fidelity reference. 
Single-fidelity fine-tuning model overcorrects towards the high-fidelity reference.
In summary, the $NPT$ ensemble highlights the same caveats identified in $NVT$, but in a more pronounced form. 
It further shows that, when models are trained only on energies, differential learning and multi-fidelity fine-tuning tend to inherit the bias of the low-fidelity reference, whereas single-fidelity fine-tuning tends to overcorrect.

\section{Discussion}
\label{s:discussion}

In summary, we investigate how well MLIPs generalize to bulk properties when trained with direct, differential, and transfer learning on high-fidelity total energies and/or forces estimated for molecular clusters. 
We utilize pretrained MLIPs as proxies to the low- and high-fidelity levels, enabling us to estimate ground-truth observables in $NVT$ and $NPT$ ensembles.
We consider both global properties, such as the total energy and density, as well as microscopic properties, including the intermolecular (O-O) and intramolecular (O--H) distributions. 
We first summarize the identified strengths and limitations of the four strategies and then discuss the best recipes for the cluster-to-bulk generalization of MLIPs. 
It should be noted, however, that the ability of an MLIP to generalise depends on how the implicit atomic body order is represented, which in turn is sensitive to the underlying architecture~\cite{chong_resolving_2025}. 
Accordingly, our conclusions are limited to ACE-based MLIPs. \\

\textbf{Direct learning:} Direct learning is feasible with only a few hundred clusters, provided both total energies and forces are included in the training set, yielding an excellent description of both global and microscopic properties. 
The main drawback of direct learning is that MLIPs are not robust to extrapolation. 
We arrive at this conclusion by performing temperature extrapolations, summarized in Fig. S9, in which $NPT$ simulations were performed up to 300\,K in 50\,K increments in the metastable solid phase.
These show that, despite excellent performance at 100\,K, direct learning on total energies and forces fails to produce stable simulations at higher temperatures. \\

\textbf{Transfer learning -- single-fidelity fine-tuning:} 
Single-fidelity fine-tuning is found to be data-efficient and robust with respect to the stability of the simulations. 
When trained on both energies and forces, this strategy yields quantitative accuracy for both global and microscopic properties in the $NVT$ and $NPT$ ensembles. 
When trained only on energies, its accuracy is sensitive to the observable and the thermodynamic ensemble. 
For the total energy and O--O distribution functions, quantitative  agreement is achieved in the $NVT$ ensemble, while higher energy RMSEs 
and skewed (overcorrected) O--O distributions are observed in the $NPT$ ensemble. 
For O--H bond distributions, both ensembles exhibit shorter (overcorrected) bond lengths. 
We also investigated whether using larger cluster sizes in the training set fixes these limitations when training only on energies. 
As summarized in Figs. S6 and S7, we do not observe systematic improvements with larger clusters: energy and force errors are saturated nearly to the same extent as for 4.5~\AA{} clusters. 
In fact, the good agreement with the density obtained for 4.5~\AA{} clusters is degraded when larger clusters are used. \\

\textbf{Differential learning:} 
We observe artifacts associated with differential learning in the context of the cluster-to-bulk generalization. 
When trained on both energies and forces, differential learning is noted to overfit to cluster data, which results in an overall poor description of global and microscopic properties and unphysical simulations with 400 clusters.
Overfitting is reduced when training only on energies, but the accuracy of the MLIPs is sensitive to the observables.
Considering global properties, the total energy RMSEs decrease by $\sim$2\,meV/atom in both ensembles, and the density error is reduced to $\sim$2\% with 1000 clusters. 
In contrast, the O--O and O--H distributions remain skewed toward the low-fidelity reference, with the deviations significantly larger in the $NPT$ ensemble than in $NVT$.
It is important to note, however, that all strategies performed poorly when trained only on energies, with differential learning yielding the best results among them.  
We also tested for the temperature extrapolation of MLIPs trained with differential learning on energies only, and note that while all simulations are stable, the errors in the predicted density increase with temperature towards the low-fidelity reference.
Thus, the results of differential learning simulations must be interpreted with caution, as stable trajectories do not necessarily imply accuracy with respect to the reference.  
A positive aspect of this approach is that its performance can be improved by increasing the cluster size in the training set. 
As shown in Fig.~S6, the description of global properties in the $NPT$ ensemble shows a modest improvement with 6~\AA{} clusters compared to 4.5~\AA{} clusters. 
For microscopic properties (Fig.~S7), the accuracy of O--H distributions improves using larger clusters, whereas the error in the O--O distributions is essentially unchanged. 
Given the systematically improvable nature of differential learning, we anticipate that data augmentation strategies could help mitigate its limitations.   \\

\textbf{Transfer learning -- multi-fidelity fine-tuning:} 
Multi-fidelity fine-tuning emerges as the most data-efficient and accurate approach for both global and microscopic properties across both ensembles when trained on energies and forces. 
As shown in Fig.~S8, multi-fidelity models also display excellent extrapolation of the density up to 300\,K, despite training clusters being representative of 100\,K only.  
There, it should be the method of choice, then the training data comprises both total energies and forces. 
Alongside these strengths, we identify two areas where multi-fidelity fine-tuning can be improved. 
First, we do not observe systematic improvements with increasing cluster size. 
When trained on 4\,\AA{} clusters comprising only five water molecules, multi-fidelity fine-tuning is the only approach that yields stable simulations with semi-quantitative agreement for the density and O--O and O--H bond lengths, as shown in Fig. S5.
In contrast, training on 6\,\AA{} clusters results in slightly worse descriptions of these observables as shown in Figs. S6 and S7.
This suggests that the fine-tuning dynamics considered here do not exhibit ``size consistency'' and may require a dedicated hyperparameter optimization per dataset. 
Second, when trained only on energies, the method performs poorly for microscopic observables, showing significant deviations in O--H bond distributions, and thus, the recipe considered here is not recommended.
Overall, multi-fidelity fine-tuning remains the most flexible approach in terms of combining low- and high-fidelity data. 
Given its demonstrated success on both clusters~\cite{cui_multi-fidelity_2025} and bulk systems~\cite{messerly_multi-fidelity_2025}, we expect that future work focused on data augmentation could yield a robust recipe for the cluster-to-bulk generalization when training on energies.

As far as the cluster-to-bulk generalization of MLIPs is concerned, our results show that this approach can be successfully applied to a range of molecular systems, provided validation is performed thoroughly. 
For an MLIP to be robust, it must accurately describe atomic forces in addition to energies. 
We reach this conclusion by noting that quantitative agreement with microscopic properties correlates strongly with low force errors, but only weakly with low energy errors or stable simulations. 
This highlights that in scenarios where forces are unavailable, it is crucial to obtain indirect probes of force accuracy, for example, by examining multiple global and microscopic observables associated with high-frequency intramolecular modes -- which, by definition, are governed by large forces. 
We further find that generalization to the $NPT$ ensemble is considerably more demanding than to $NVT$, making careful validation essential. 
Finally, the strong correlation between accurate microscopic properties and low-force errors across trajectory ensembles motivates two key directions for future work:  
(i) developing MLIP strategies capable of accurate force prediction when training only on energies, and  
(ii) advancing efficient, mainstream implementations of gradients~\cite{pinskiAnalyticalGradientDomainbased2019,zhangPerformantAutomaticDifferentiation2024,slootmanAccurateQuantumMonte2024} for cWFT methods, both for clusters (training) and bulk phases (validation), to enable robust and cost-effective cluster-to-bulk generalization.

\section*{Acknowledgments}
We thank Michele Ceriotti, Ilyes Batatia, and Sander Vandenhaute for their valuable comments on the manuscript.
MJG acknowledges the UCL Department of Physics and Astronomy's PhD studentship, and VK acknowledges support from UCL's startup funds.
ML further acknowledges the CCP summer bursary.
The Flatiron Institute is a division of the Simons Foundation.
We are grateful for computational support from the Swiss National Supercomputing Centre (CSCS) under project s1288 on Alps and UCL Myriad High Performance Computing Facility (Myriad@UCL). This work used computing equipment funded by the Research Capital Investment Fund (RCIF) provided by UKRI, and partially funded by the UCL Cosmoparticle Initiative.

\section*{Author declarations}

\subsection*{Conflict of Interest}
The authors have no conflicts to disclose.

\section*{Data availability}
All the training data, code, and scripts required to reproduce the results will be made available upon publication.

\section*{References}



\section*{Supplementary material}

**CONTENTS**

# S1. REFERENCE OBSERVABLES AT LOW- AND HIGH-FIDELITY

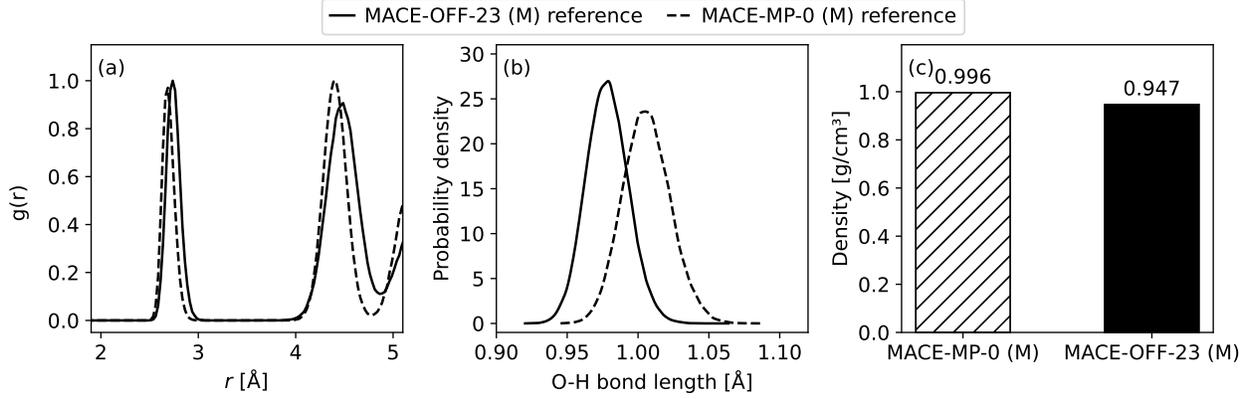

FIG. S1. Comparison of microscopic properties from $NPT$ simulations of hexagonal ice using the medium (M) MACE-MP-0 and MACE-OFF-23 models. (a) O-O pair correlation functions, (b) distributions of O-H bond lengths, and (c) average densities from the $NPT$ trajectories.

# S2. VALIDATION ERRORS

## A. Energy and force RMSEs

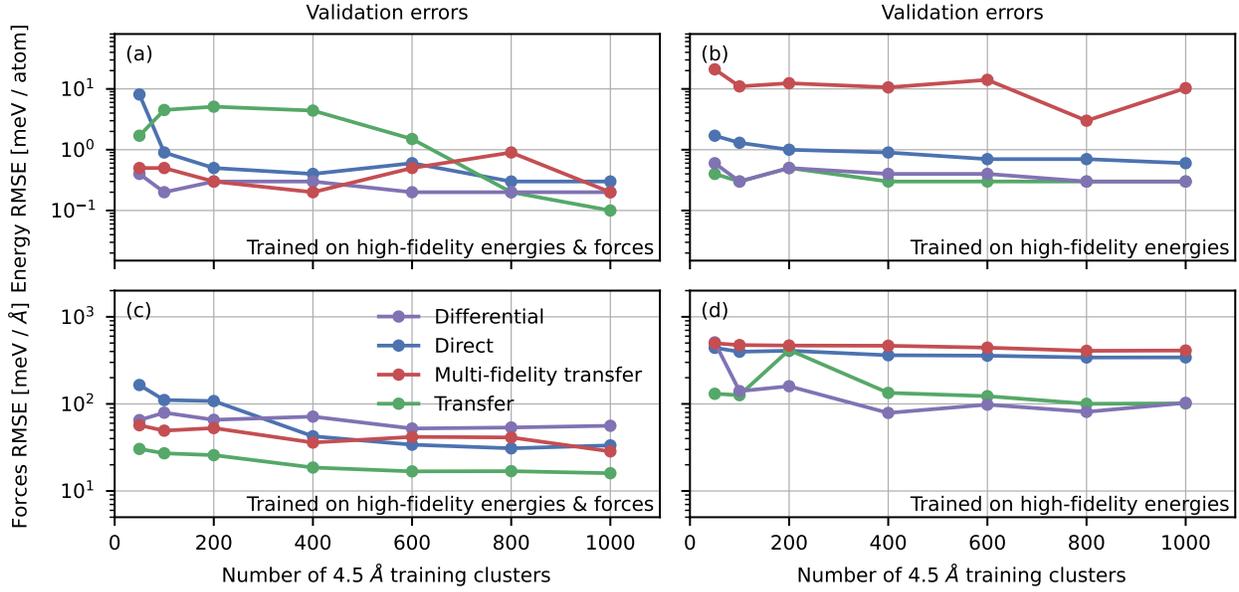

FIG. S2. RMSEs for ice-Ih on the validation set as a function of the number of 4.5 Å clusters in the training set. Panels (a) and (c) show the energy-per-atom and mean force RMSEs, respectively, for models trained on both high-fidelity energies and forces. Panels (b) and (d) show the corresponding results for models trained on high-fidelity energies only.

## B. Hyperparameter optimization

| Training protocol | $(E_{\mathrm{RMSE}}^{4.5}, F_{\mathrm{RMSE}}^{4.5})$ | $(\lambda_U^{4.5}, \lambda_{\mathbf{f}}^{4.5})$ | $(E_{\mathrm{RMSE}}^{4.5}, F_{\mathrm{RMSE}}^{4.5})$ | $(\lambda_U^{4.5}, \lambda_{\mathbf{f}}^{4.5})$ |
|---|---|---|---|---|
| Direct | (0.3, 33.4) | (1000, 10) | (0.6, 342.2) | (1000, 0) |
| Differential | (0.2, 56.1) | (1000, 1) | (0.3,102.8) | (1000,0) |
| Transfer | (0.1, 16) | (10, 10) | (0.3,101.3) | (100,0) |
| Multi-fidelity transfer | (0.2, 28.5) | (100, 10) | (0.3, 250.9) | (10, 0) |

TABLE S1. Hyperparameter optimization results for energy and force errors, as well as the optimal training weights, across four training strategies. $(E_{\mathrm{RMSE}}^{4.5}, F_{\mathrm{RMSE}}^{4.5})$ denote the energy and force RMSEs obtained when training on clusters of size 4.5 Å. $(\lambda_U^i, \lambda_{\mathbf{f}}^i)$ denote the corresponding optimal energy and force weights when training on 4.5 Å clusters.

## S3. VALIDATION WITH DFT DATA

The high-fidelity DFT calculations were revPBE-D3, with zero-damping in the D3 dispersion, and the low-fidelity was a medium MACE-MP-0 model with the pretrained head set to the MPTrj dataset. The reference bulk density was calculated from periodic calculations in the Vienna *ab initio* Simulations Package (VASP)[1,2]. We used hard projector augmented wave (PAW) potentials (version 54) and an energy cutoff of 1000 eV on the $\Gamma$-point. The cluster calculations were performed in ORCA v5.0.3[3] for the def2-QZVPPD basis set. The isolated atomic energies were $U_H$ = -13.72710738642251 eV and $U_O$ = -2042.922011881597 eV for the revPBE-D3 data.

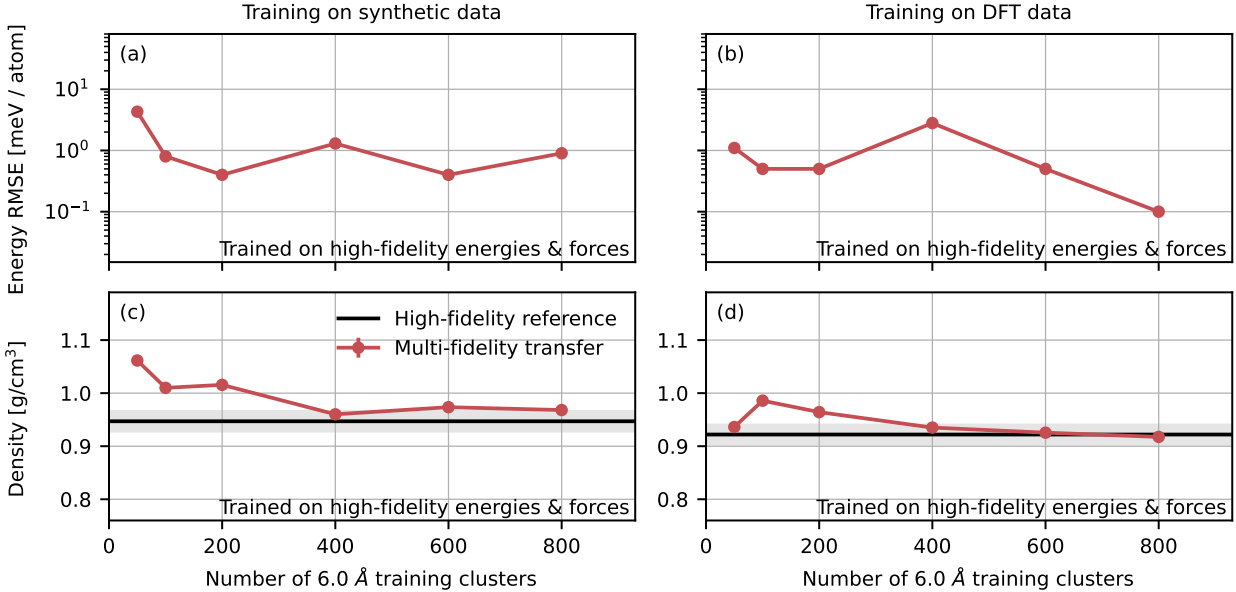

FIG. S3. Validation energy RMSEs and average densities of $NPT$ trajectories for models trained on 6.0 Å clusters: panels (a,c) show results using synthetic data, and panels (b,d) show results using DFT data. 'High-fidelity reference' in panel (c) corresponds to the MACE-OFF-23 reference density, whereas the high-fidelity reference in panel (d) is the revPBE-D3 reference density.

# S4. FORCE RMSE IN THE *NVT* AND *NPT* ENSEMBLES

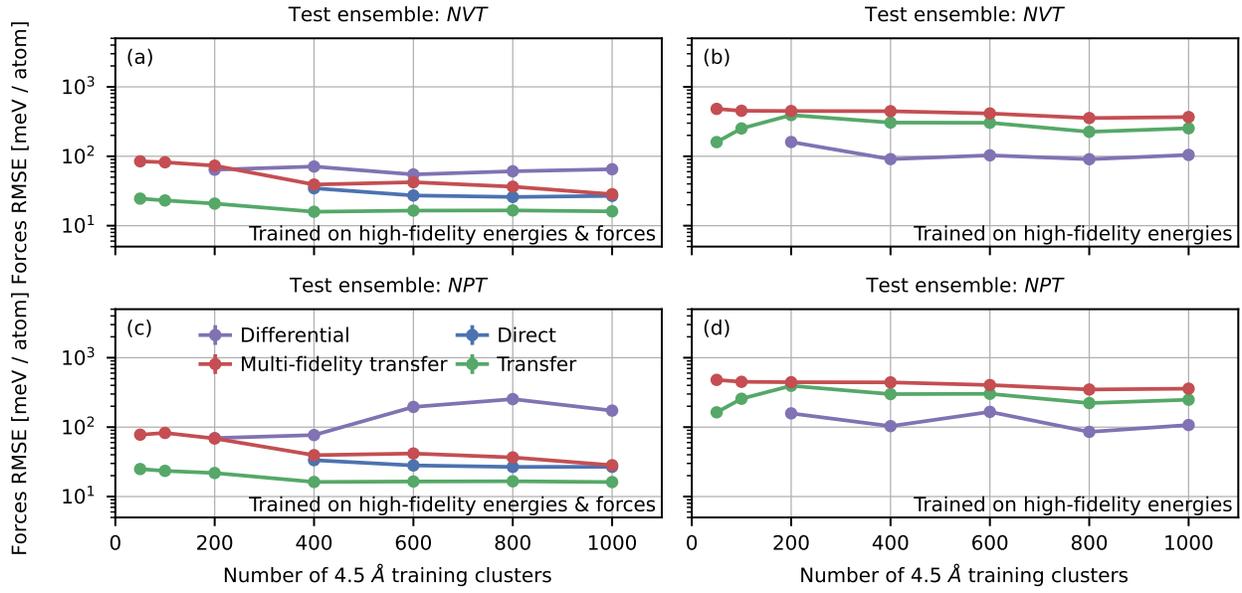

FIG. S4. Force RMSEs under different training conditions: (a,c) trained on high-fidelity energies and forces; (b,d) trained on high-fidelity energies only. Results are shown for testing in the *NVT* (a,b) and *NPT* (c,d) ensembles.

# S5. TRAINING ON 4 Å AND 6 Å CLUSTERS

## A. 4 Å clusters

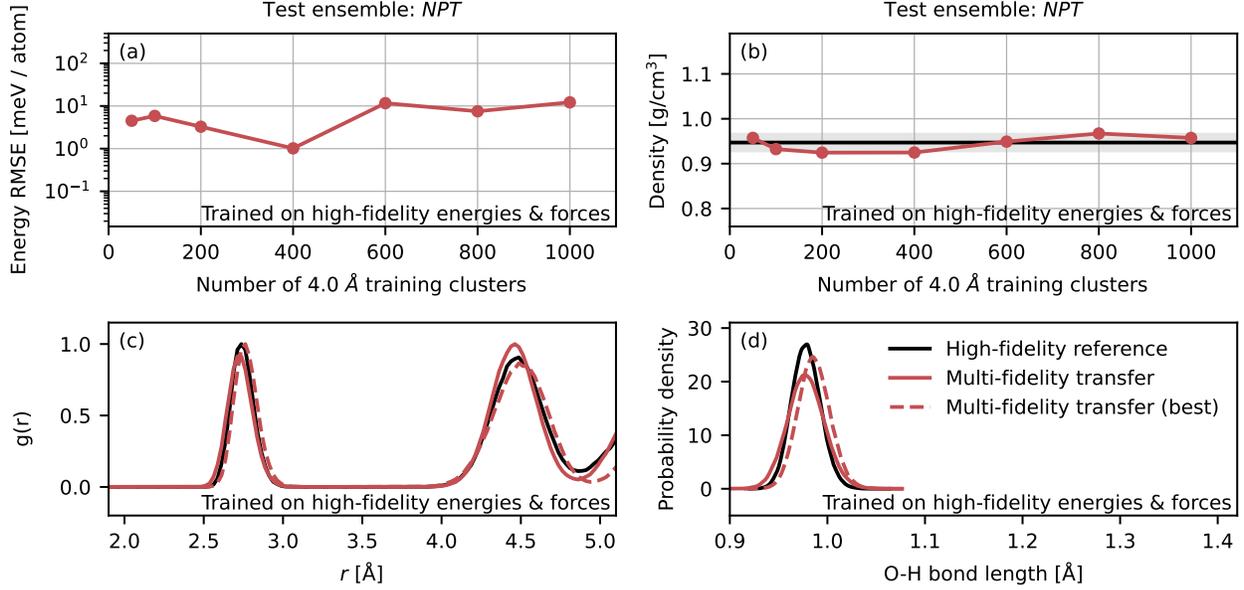

FIG. S5. (a) Energy RMSEs, (b) densities of $NPT$ trajectories, (c) O-O pair correlation functions, and (d) O-H bond distributions obtained with the multi-fidelity transfer learning approach trained on 4.0 Å clusters. 'Multi-fidelity transfer (best)' refers to the multi-fidelity transfer model that achieved the lowest energy RMSE and was trained on 400 clusters.

## B. 6 Å clusters

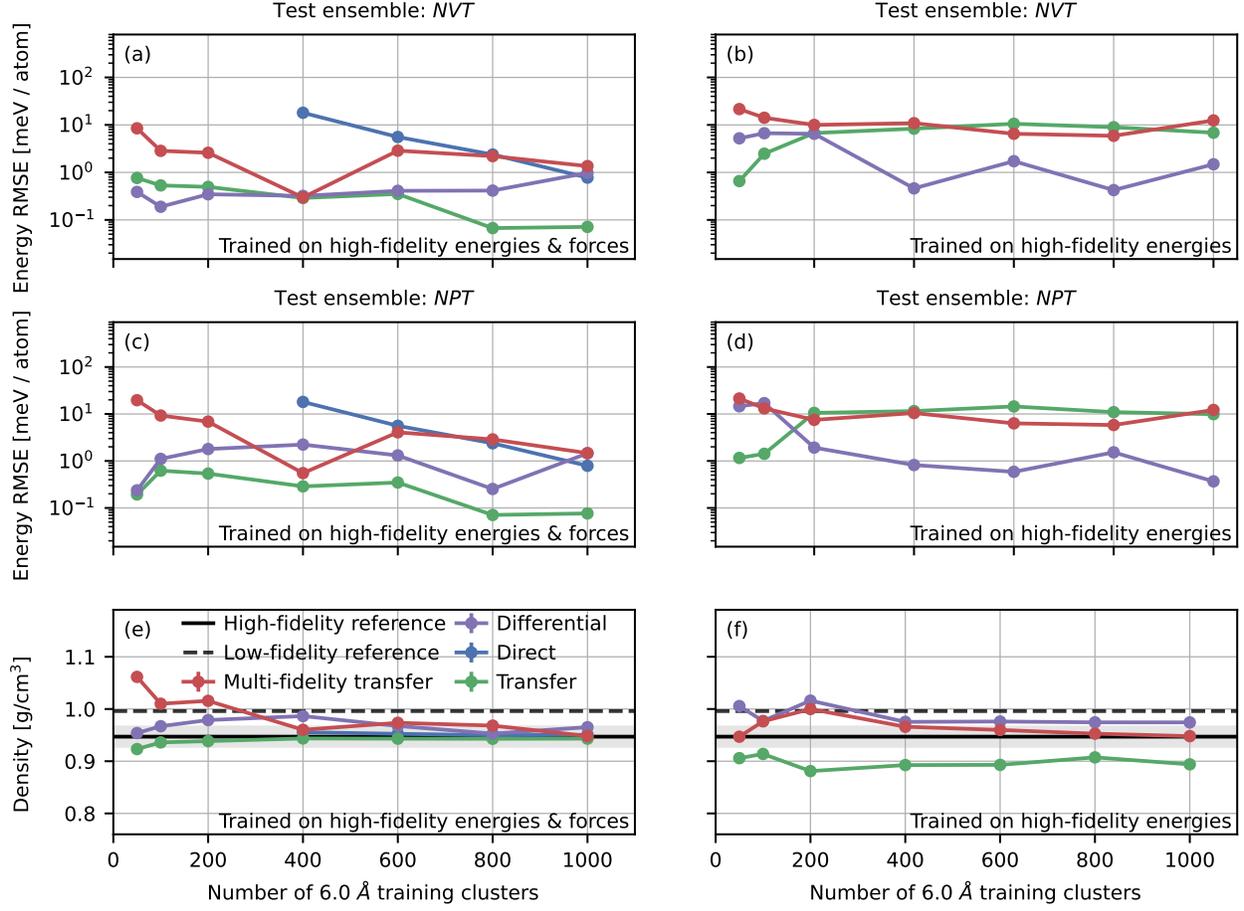

FIG. S6. Energy RMSEs and average densities from $NPT$ simulations for models trained on 6.0 Å clusters. Left-hand panels show results for models trained on both high-fidelity energies and forces, tested in the $NVT$ (a) and $NPT$ (c,e) ensembles. Right-hand panels show results for models trained on high-fidelity energies only, tested in the $NVT$ (b) and $NPT$ (d,f) ensembles.

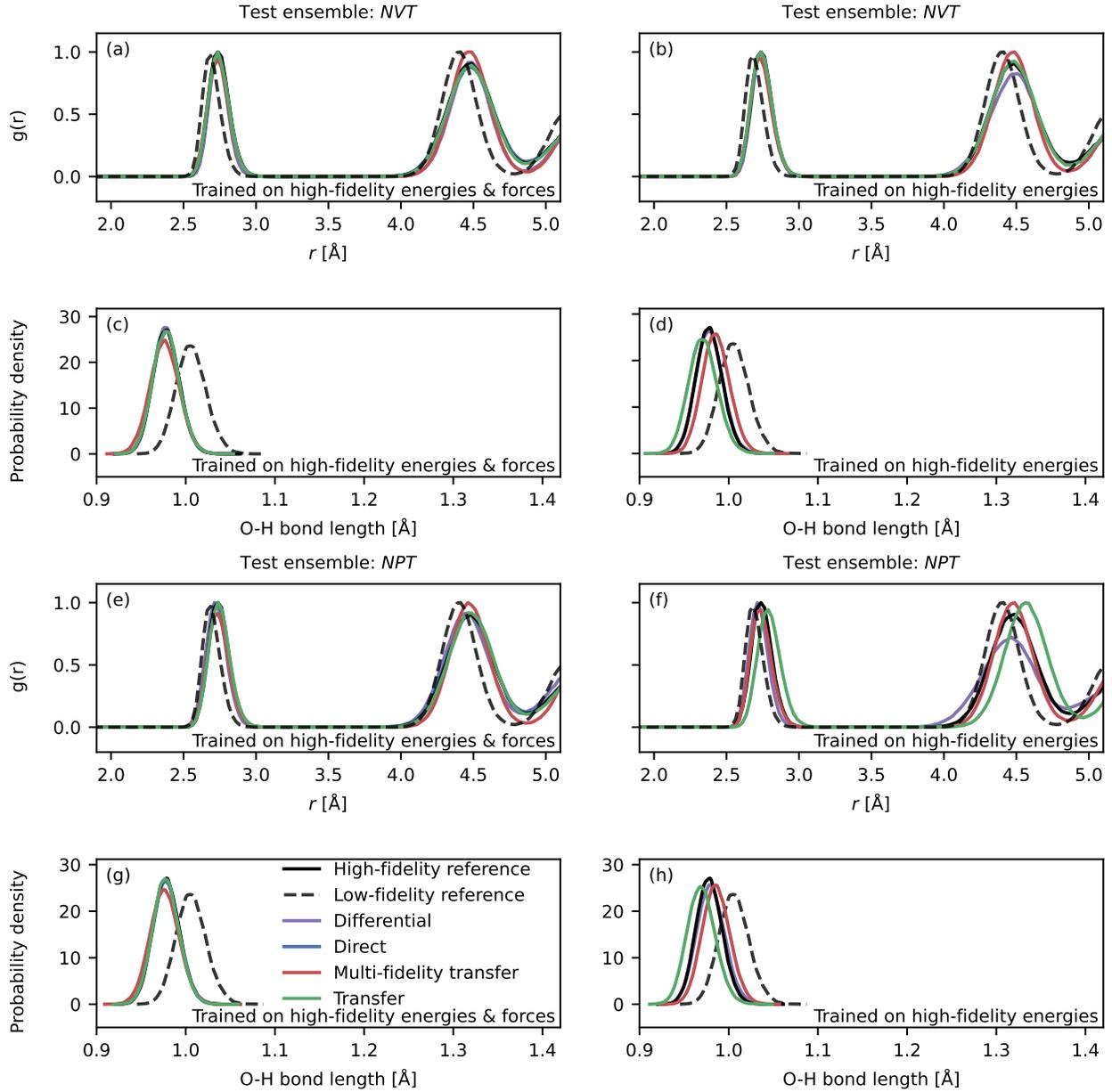

FIG. S7. Structural properties from $NPT$ simulations using models trained on 6.0 Å clusters. Left-hand panels show results for models trained on both high-fidelity energies and forces, tested in the $NVT$ (a,c) and $NPT$ (e,g) ensembles. Right-hand panels show results for models trained on high-fidelity energies only, tested in the $NVT$ (b,d) and $NPT$ (f,h) ensembles.

# S6. TEMPERATURE EXTRAPOLATIONS

## A. Density

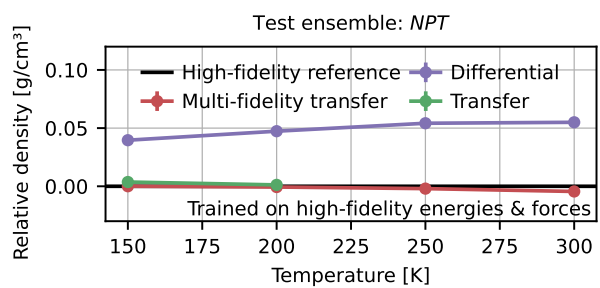

FIG. S8. Relative densities of structures sampled from *NPT* simulations at four temperatures shown for different training methods relative to the high-fidelity reference values.

## B. Microscopic observables

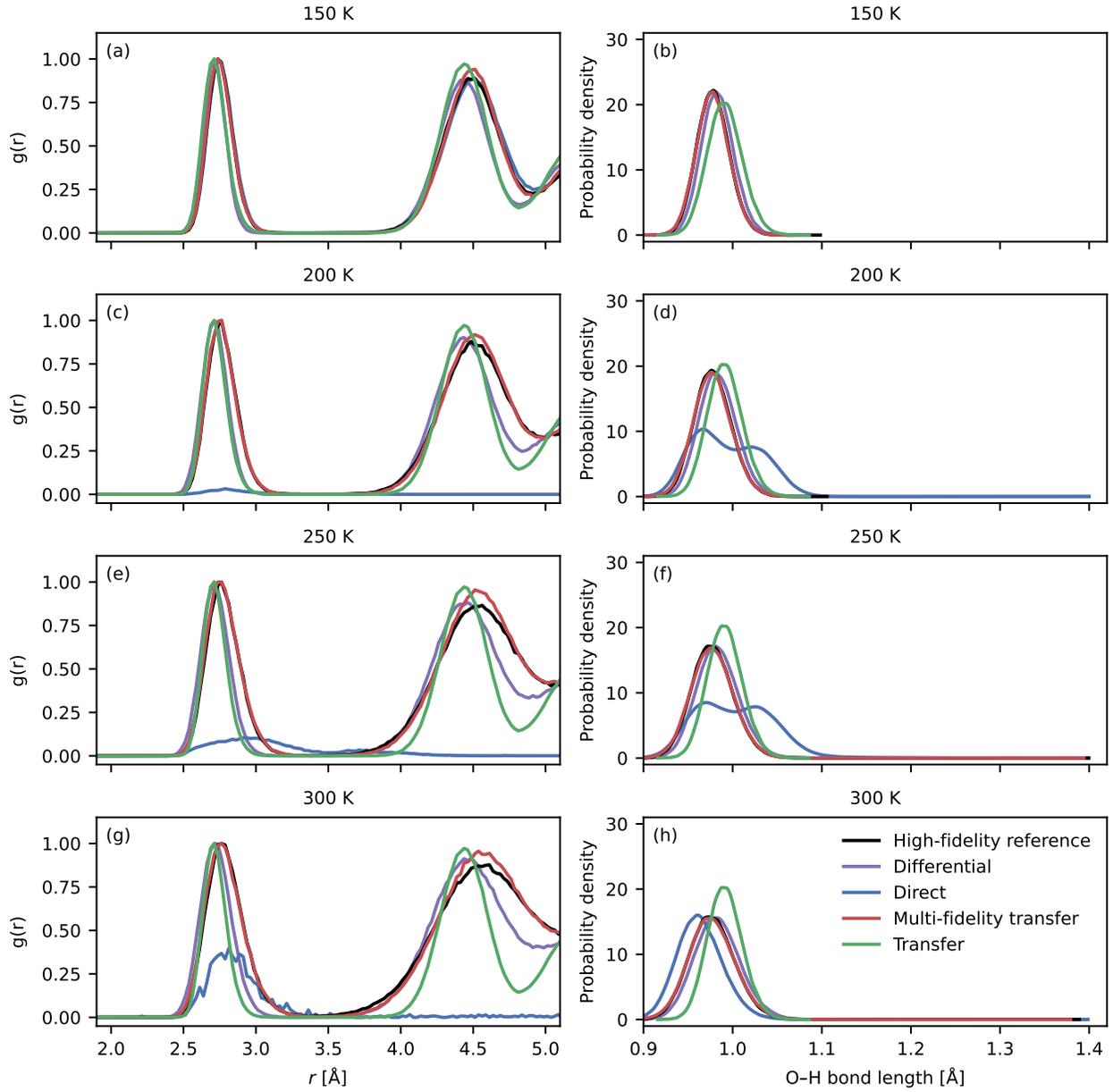

FIG. S9. O-O pair correlation functions (a,c,e,g) and O-H bond-length distributions (b,d,f,h) of *NPT* trajectories for the temperature extrapolation simulations.

## REFERENCES

[1] G. Kresse and J. Furthmüller, Computational Materials Science **6**, 15 (1996).

[2] G. Kresse and J. Furthmüller, Physical Review B **54**, 11169 (1996).


\begin{thebibliography}{62}%
\makeatletter
\providecommand \@ifxundefined [1]{%
 \@ifx{#1\undefined}
}%
\providecommand \@ifnum [1]{%
 \ifnum #1\expandafter \@firstoftwo
 \else \expandafter \@secondoftwo
 \fi
}%
\providecommand \@ifx [1]{%
 \ifx #1\expandafter \@firstoftwo
 \else \expandafter \@secondoftwo
 \fi
}%
\providecommand \natexlab [1]{#1}%
\providecommand \enquote  [1]{``#1''}%
\providecommand \bibnamefont  [1]{#1}%
\providecommand \bibfnamefont [1]{#1}%
\providecommand \citenamefont [1]{#1}%
\providecommand \href@noop [0]{\@secondoftwo}%
\providecommand \href [0]{\begingroup \@sanitize@url \@href}%
\providecommand \@href[1]{\@@startlink{#1}\@@href}%
\providecommand \@@href[1]{\endgroup#1\@@endlink}%
\providecommand \@sanitize@url [0]{\catcode `\\12\catcode `\$12\catcode
  `\&12\catcode `\#12\catcode `\^12\catcode `\_12\catcode `\%12\relax}%
\providecommand \@@startlink[1]{}%
\providecommand \@@endlink[0]{}%
\providecommand \url  [0]{\begingroup\@sanitize@url \@url }%
\providecommand \@url [1]{\endgroup\@href {#1}{\urlprefix }}%
\providecommand \urlprefix  [0]{URL }%
\providecommand \Eprint [0]{\href }%
\providecommand \doibase [0]{http://dx.doi.org/}%
\providecommand \selectlanguage [0]{\@gobble}%
\providecommand \bibinfo  [0]{\@secondoftwo}%
\providecommand \bibfield  [0]{\@secondoftwo}%
\providecommand \translation [1]{[#1]}%
\providecommand \BibitemOpen [0]{}%
\providecommand \bibitemStop [0]{}%
\providecommand \bibitemNoStop [0]{.\EOS\space}%
\providecommand \EOS [0]{\spacefactor3000\relax}%
\providecommand \BibitemShut  [1]{\csname bibitem#1\endcsname}%
\let\auto@bib@innerbib\@empty
\bibitem [{\citenamefont {Kovács}\ \emph {et~al.}(2025)\citenamefont
  {Kovács}, \citenamefont {Moore}, \citenamefont {Browning}, \citenamefont
  {Batatia}, \citenamefont {Horton}, \citenamefont {Pu}, \citenamefont {Kapil},
  \citenamefont {Witt}, \citenamefont {Magdău}, \citenamefont {Cole},\ and\
  \citenamefont {Csányi}}]{kovacs_mace-off_2025}%
  \BibitemOpen
  \bibfield  {author} {\bibinfo {author} {\bibfnamefont {D.~P.}\ \bibnamefont
  {Kovács}}, \bibinfo {author} {\bibfnamefont {J.~H.}\ \bibnamefont {Moore}},
  \bibinfo {author} {\bibfnamefont {N.~J.}\ \bibnamefont {Browning}}, \bibinfo
  {author} {\bibfnamefont {I.}~\bibnamefont {Batatia}}, \bibinfo {author}
  {\bibfnamefont {J.~T.}\ \bibnamefont {Horton}}, \bibinfo {author}
  {\bibfnamefont {Y.}~\bibnamefont {Pu}}, \bibinfo {author} {\bibfnamefont
  {V.}~\bibnamefont {Kapil}}, \bibinfo {author} {\bibfnamefont {W.~C.}\
  \bibnamefont {Witt}}, \bibinfo {author} {\bibfnamefont {I.-B.}\ \bibnamefont
  {Magdău}}, \bibinfo {author} {\bibfnamefont {D.~J.}\ \bibnamefont {Cole}}, \
  and\ \bibinfo {author} {\bibfnamefont {G.}~\bibnamefont {Csányi}},\ }\href
  {\doibase 10.1021/jacs.4c07099} {\bibfield  {journal} {\bibinfo  {journal}
  {Journal of the American Chemical Society}\ }\textbf {\bibinfo {volume}
  {147}},\ \bibinfo {pages} {17598} (\bibinfo {year} {2025})}\BibitemShut
  {NoStop}%
\bibitem [{\citenamefont {Batatia}\ \emph {et~al.}(2024)\citenamefont
  {Batatia}, \citenamefont {Benner}, \citenamefont {Chiang}, \citenamefont
  {Elena}, \citenamefont {Kovács}, \citenamefont {Riebesell}, \citenamefont
  {Advincula}, \citenamefont {Asta}, \citenamefont {Avaylon}, \citenamefont
  {Baldwin}, \citenamefont {Berger}, \citenamefont {Bernstein}, \citenamefont
  {Bhowmik}, \citenamefont {Blau}, \citenamefont {Cărare}, \citenamefont
  {Darby}, \citenamefont {De}, \citenamefont {Della~Pia}, \citenamefont
  {Deringer}, \citenamefont {Elijošius}, \citenamefont {El-Machachi},
  \citenamefont {Falcioni}, \citenamefont {Fako}, \citenamefont {Ferrari},
  \citenamefont {Genreith-Schriever}, \citenamefont {George}, \citenamefont
  {Goodall}, \citenamefont {Grey}, \citenamefont {Grigorev}, \citenamefont
  {Han}, \citenamefont {Handley}, \citenamefont {Heenen}, \citenamefont
  {Hermansson}, \citenamefont {Holm}, \citenamefont {Jaafar}, \citenamefont
  {Hofmann}, \citenamefont {Jakob}, \citenamefont {Jung}, \citenamefont
  {Kapil}, \citenamefont {Kaplan}, \citenamefont {Karimitari}, \citenamefont
  {Kermode}, \citenamefont {Kroupa}, \citenamefont {Kullgren}, \citenamefont
  {Kuner}, \citenamefont {Kuryla}, \citenamefont {Liepuoniute}, \citenamefont
  {Margraf}, \citenamefont {Magdău}, \citenamefont {Michaelides},
  \citenamefont {Moore}, \citenamefont {Naik}, \citenamefont {Niblett},
  \citenamefont {Norwood}, \citenamefont {O'Neill}, \citenamefont {Ortner},
  \citenamefont {Persson}, \citenamefont {Reuter}, \citenamefont {Rosen},
  \citenamefont {Schaaf}, \citenamefont {Schran}, \citenamefont {Shi},
  \citenamefont {Sivonxay}, \citenamefont {Stenczel}, \citenamefont {Svahn},
  \citenamefont {Sutton}, \citenamefont {Swinburne}, \citenamefont {Tilly},
  \citenamefont {van~der Oord}, \citenamefont {Varga-Umbrich}, \citenamefont
  {Vegge}, \citenamefont {Vondrák}, \citenamefont {Wang}, \citenamefont
  {Witt}, \citenamefont {Zills},\ and\ \citenamefont
  {Csányi}}]{batatia_foundation_2024}%
  \BibitemOpen
  \bibfield  {author} {\bibinfo {author} {\bibfnamefont {I.}~\bibnamefont
  {Batatia}}, \bibinfo {author} {\bibfnamefont {P.}~\bibnamefont {Benner}},
  \bibinfo {author} {\bibfnamefont {Y.}~\bibnamefont {Chiang}}, \bibinfo
  {author} {\bibfnamefont {A.~M.}\ \bibnamefont {Elena}}, \bibinfo {author}
  {\bibfnamefont {D.~P.}\ \bibnamefont {Kovács}}, \bibinfo {author}
  {\bibfnamefont {J.}~\bibnamefont {Riebesell}}, \bibinfo {author}
  {\bibfnamefont {X.~R.}\ \bibnamefont {Advincula}}, \bibinfo {author}
  {\bibfnamefont {M.}~\bibnamefont {Asta}}, \bibinfo {author} {\bibfnamefont
  {M.}~\bibnamefont {Avaylon}}, \bibinfo {author} {\bibfnamefont {W.~J.}\
  \bibnamefont {Baldwin}}, \bibinfo {author} {\bibfnamefont {F.}~\bibnamefont
  {Berger}}, \bibinfo {author} {\bibfnamefont {N.}~\bibnamefont {Bernstein}},
  \bibinfo {author} {\bibfnamefont {A.}~\bibnamefont {Bhowmik}}, \bibinfo
  {author} {\bibfnamefont {S.~M.}\ \bibnamefont {Blau}}, \bibinfo {author}
  {\bibfnamefont {V.}~\bibnamefont {Cărare}}, \bibinfo {author} {\bibfnamefont
  {J.~P.}\ \bibnamefont {Darby}}, \bibinfo {author} {\bibfnamefont
  {S.}~\bibnamefont {De}}, \bibinfo {author} {\bibfnamefont {F.}~\bibnamefont
  {Della~Pia}}, \bibinfo {author} {\bibfnamefont {V.~L.}\ \bibnamefont
  {Deringer}}, \bibinfo {author} {\bibfnamefont {R.}~\bibnamefont
  {Elijošius}}, \bibinfo {author} {\bibfnamefont {Z.}~\bibnamefont
  {El-Machachi}}, \bibinfo {author} {\bibfnamefont {F.}~\bibnamefont
  {Falcioni}}, \bibinfo {author} {\bibfnamefont {E.}~\bibnamefont {Fako}},
  \bibinfo {author} {\bibfnamefont {A.~C.}\ \bibnamefont {Ferrari}}, \bibinfo
  {author} {\bibfnamefont {A.}~\bibnamefont {Genreith-Schriever}}, \bibinfo
  {author} {\bibfnamefont {J.}~\bibnamefont {George}}, \bibinfo {author}
  {\bibfnamefont {R.~E.~A.}\ \bibnamefont {Goodall}}, \bibinfo {author}
  {\bibfnamefont {C.~P.}\ \bibnamefont {Grey}}, \bibinfo {author}
  {\bibfnamefont {P.}~\bibnamefont {Grigorev}}, \bibinfo {author}
  {\bibfnamefont {S.}~\bibnamefont {Han}}, \bibinfo {author} {\bibfnamefont
  {W.}~\bibnamefont {Handley}}, \bibinfo {author} {\bibfnamefont {H.~H.}\
  \bibnamefont {Heenen}}, \bibinfo {author} {\bibfnamefont {K.}~\bibnamefont
  {Hermansson}}, \bibinfo {author} {\bibfnamefont {C.}~\bibnamefont {Holm}},
  \bibinfo {author} {\bibfnamefont {J.}~\bibnamefont {Jaafar}}, \bibinfo
  {author} {\bibfnamefont {S.}~\bibnamefont {Hofmann}}, \bibinfo {author}
  {\bibfnamefont {K.~S.}\ \bibnamefont {Jakob}}, \bibinfo {author}
  {\bibfnamefont {H.}~\bibnamefont {Jung}}, \bibinfo {author} {\bibfnamefont
  {V.}~\bibnamefont {Kapil}}, \bibinfo {author} {\bibfnamefont {A.~D.}\
  \bibnamefont {Kaplan}}, \bibinfo {author} {\bibfnamefont {N.}~\bibnamefont
  {Karimitari}}, \bibinfo {author} {\bibfnamefont {J.~R.}\ \bibnamefont
  {Kermode}}, \bibinfo {author} {\bibfnamefont {N.}~\bibnamefont {Kroupa}},
  \bibinfo {author} {\bibfnamefont {J.}~\bibnamefont {Kullgren}}, \bibinfo
  {author} {\bibfnamefont {M.~C.}\ \bibnamefont {Kuner}}, \bibinfo {author}
  {\bibfnamefont {D.}~\bibnamefont {Kuryla}}, \bibinfo {author} {\bibfnamefont
  {G.}~\bibnamefont {Liepuoniute}}, \bibinfo {author} {\bibfnamefont {J.~T.}\
  \bibnamefont {Margraf}}, \bibinfo {author} {\bibfnamefont {I.-B.}\
  \bibnamefont {Magdău}}, \bibinfo {author} {\bibfnamefont {A.}~\bibnamefont
  {Michaelides}}, \bibinfo {author} {\bibfnamefont {J.~H.}\ \bibnamefont
  {Moore}}, \bibinfo {author} {\bibfnamefont {A.~A.}\ \bibnamefont {Naik}},
  \bibinfo {author} {\bibfnamefont {S.~P.}\ \bibnamefont {Niblett}}, \bibinfo
  {author} {\bibfnamefont {S.~W.}\ \bibnamefont {Norwood}}, \bibinfo {author}
  {\bibfnamefont {N.}~\bibnamefont {O'Neill}}, \bibinfo {author} {\bibfnamefont
  {C.}~\bibnamefont {Ortner}}, \bibinfo {author} {\bibfnamefont {K.~A.}\
  \bibnamefont {Persson}}, \bibinfo {author} {\bibfnamefont {K.}~\bibnamefont
  {Reuter}}, \bibinfo {author} {\bibfnamefont {A.~S.}\ \bibnamefont {Rosen}},
  \bibinfo {author} {\bibfnamefont {L.~L.}\ \bibnamefont {Schaaf}}, \bibinfo
  {author} {\bibfnamefont {C.}~\bibnamefont {Schran}}, \bibinfo {author}
  {\bibfnamefont {B.~X.}\ \bibnamefont {Shi}}, \bibinfo {author} {\bibfnamefont
  {E.}~\bibnamefont {Sivonxay}}, \bibinfo {author} {\bibfnamefont {T.~K.}\
  \bibnamefont {Stenczel}}, \bibinfo {author} {\bibfnamefont {V.}~\bibnamefont
  {Svahn}}, \bibinfo {author} {\bibfnamefont {C.}~\bibnamefont {Sutton}},
  \bibinfo {author} {\bibfnamefont {T.~D.}\ \bibnamefont {Swinburne}}, \bibinfo
  {author} {\bibfnamefont {J.}~\bibnamefont {Tilly}}, \bibinfo {author}
  {\bibfnamefont {C.}~\bibnamefont {van~der Oord}}, \bibinfo {author}
  {\bibfnamefont {E.}~\bibnamefont {Varga-Umbrich}}, \bibinfo {author}
  {\bibfnamefont {T.}~\bibnamefont {Vegge}}, \bibinfo {author} {\bibfnamefont
  {M.}~\bibnamefont {Vondrák}}, \bibinfo {author} {\bibfnamefont
  {Y.}~\bibnamefont {Wang}}, \bibinfo {author} {\bibfnamefont {W.~C.}\
  \bibnamefont {Witt}}, \bibinfo {author} {\bibfnamefont {F.}~\bibnamefont
  {Zills}}, \ and\ \bibinfo {author} {\bibfnamefont {G.}~\bibnamefont
  {Csányi}},\ }\href {\doibase 10.48550/arXiv.2401.00096} {\enquote {\bibinfo
  {title} {A foundation model for atomistic materials chemistry},}\ } (\bibinfo
  {year} {2024}),\ \bibinfo {note} {arXiv:2401.00096 [cond-mat,
  physics:physics]}\BibitemShut {NoStop}%
\bibitem [{\citenamefont {Merchant}\ \emph {et~al.}(2023)\citenamefont
  {Merchant}, \citenamefont {Batzner}, \citenamefont {Schoenholz},
  \citenamefont {Aykol}, \citenamefont {Cheon},\ and\ \citenamefont
  {Cubuk}}]{merchant_scaling_2023}%
  \BibitemOpen
  \bibfield  {author} {\bibinfo {author} {\bibfnamefont {A.}~\bibnamefont
  {Merchant}}, \bibinfo {author} {\bibfnamefont {S.}~\bibnamefont {Batzner}},
  \bibinfo {author} {\bibfnamefont {S.~S.}\ \bibnamefont {Schoenholz}},
  \bibinfo {author} {\bibfnamefont {M.}~\bibnamefont {Aykol}}, \bibinfo
  {author} {\bibfnamefont {G.}~\bibnamefont {Cheon}}, \ and\ \bibinfo {author}
  {\bibfnamefont {E.~D.}\ \bibnamefont {Cubuk}},\ }\href {\doibase
  10.1038/s41586-023-06735-9} {\bibfield  {journal} {\bibinfo  {journal}
  {Nature}\ }\textbf {\bibinfo {volume} {624}},\ \bibinfo {pages} {80}
  (\bibinfo {year} {2023})}\BibitemShut {NoStop}%
\bibitem [{\citenamefont {Yang}\ \emph {et~al.}(2024)\citenamefont {Yang},
  \citenamefont {Hu}, \citenamefont {Zhou}, \citenamefont {Liu}, \citenamefont
  {Shi}, \citenamefont {Li}, \citenamefont {Li}, \citenamefont {Chen},
  \citenamefont {Chen}, \citenamefont {Zeni}, \citenamefont {Horton},
  \citenamefont {Pinsler}, \citenamefont {Fowler}, \citenamefont {Zügner},
  \citenamefont {Xie}, \citenamefont {Smith}, \citenamefont {Sun},
  \citenamefont {Wang}, \citenamefont {Kong}, \citenamefont {Liu},
  \citenamefont {Hao},\ and\ \citenamefont {Lu}}]{yang_mattersim_2024}%
  \BibitemOpen
  \bibfield  {author} {\bibinfo {author} {\bibfnamefont {H.}~\bibnamefont
  {Yang}}, \bibinfo {author} {\bibfnamefont {C.}~\bibnamefont {Hu}}, \bibinfo
  {author} {\bibfnamefont {Y.}~\bibnamefont {Zhou}}, \bibinfo {author}
  {\bibfnamefont {X.}~\bibnamefont {Liu}}, \bibinfo {author} {\bibfnamefont
  {Y.}~\bibnamefont {Shi}}, \bibinfo {author} {\bibfnamefont {J.}~\bibnamefont
  {Li}}, \bibinfo {author} {\bibfnamefont {G.}~\bibnamefont {Li}}, \bibinfo
  {author} {\bibfnamefont {Z.}~\bibnamefont {Chen}}, \bibinfo {author}
  {\bibfnamefont {S.}~\bibnamefont {Chen}}, \bibinfo {author} {\bibfnamefont
  {C.}~\bibnamefont {Zeni}}, \bibinfo {author} {\bibfnamefont {M.}~\bibnamefont
  {Horton}}, \bibinfo {author} {\bibfnamefont {R.}~\bibnamefont {Pinsler}},
  \bibinfo {author} {\bibfnamefont {A.}~\bibnamefont {Fowler}}, \bibinfo
  {author} {\bibfnamefont {D.}~\bibnamefont {Zügner}}, \bibinfo {author}
  {\bibfnamefont {T.}~\bibnamefont {Xie}}, \bibinfo {author} {\bibfnamefont
  {J.}~\bibnamefont {Smith}}, \bibinfo {author} {\bibfnamefont
  {L.}~\bibnamefont {Sun}}, \bibinfo {author} {\bibfnamefont {Q.}~\bibnamefont
  {Wang}}, \bibinfo {author} {\bibfnamefont {L.}~\bibnamefont {Kong}}, \bibinfo
  {author} {\bibfnamefont {C.}~\bibnamefont {Liu}}, \bibinfo {author}
  {\bibfnamefont {H.}~\bibnamefont {Hao}}, \ and\ \bibinfo {author}
  {\bibfnamefont {Z.}~\bibnamefont {Lu}},\ }\href {\doibase
  10.48550/arXiv.2405.04967} {\enquote {\bibinfo {title} {{MatterSim}: {A}
  {Deep} {Learning} {Atomistic} {Model} {Across} {Elements}, {Temperatures} and
  {Pressures}},}\ } (\bibinfo {year} {2024}),\ \bibinfo {note}
  {arXiv:2405.04967 [cond-mat]}\BibitemShut {NoStop}%
\bibitem [{\citenamefont {Unke}\ \emph {et~al.}(2024)\citenamefont {Unke},
  \citenamefont {Stöhr}, \citenamefont {Ganscha}, \citenamefont {Unterthiner},
  \citenamefont {Maennel}, \citenamefont {Kashubin}, \citenamefont {Ahlin},
  \citenamefont {Gastegger}, \citenamefont {Medrano~Sandonas}, \citenamefont
  {Berryman}, \citenamefont {Tkatchenko},\ and\ \citenamefont
  {Müller}}]{unke_biomolecular_2024}%
  \BibitemOpen
  \bibfield  {author} {\bibinfo {author} {\bibfnamefont {O.~T.}\ \bibnamefont
  {Unke}}, \bibinfo {author} {\bibfnamefont {M.}~\bibnamefont {Stöhr}},
  \bibinfo {author} {\bibfnamefont {S.}~\bibnamefont {Ganscha}}, \bibinfo
  {author} {\bibfnamefont {T.}~\bibnamefont {Unterthiner}}, \bibinfo {author}
  {\bibfnamefont {H.}~\bibnamefont {Maennel}}, \bibinfo {author} {\bibfnamefont
  {S.}~\bibnamefont {Kashubin}}, \bibinfo {author} {\bibfnamefont
  {D.}~\bibnamefont {Ahlin}}, \bibinfo {author} {\bibfnamefont
  {M.}~\bibnamefont {Gastegger}}, \bibinfo {author} {\bibfnamefont
  {L.}~\bibnamefont {Medrano~Sandonas}}, \bibinfo {author} {\bibfnamefont
  {J.~T.}\ \bibnamefont {Berryman}}, \bibinfo {author} {\bibfnamefont
  {A.}~\bibnamefont {Tkatchenko}}, \ and\ \bibinfo {author} {\bibfnamefont
  {K.-R.}\ \bibnamefont {Müller}},\ }\href {\doibase 10.1126/sciadv.adn4397}
  {\bibfield  {journal} {\bibinfo  {journal} {Science Advances}\ }\textbf
  {\bibinfo {volume} {10}},\ \bibinfo {pages} {eadn4397} (\bibinfo {year}
  {2024})}\BibitemShut {NoStop}%
\bibitem [{\citenamefont {Feibelman}\ \emph {et~al.}(2001)\citenamefont
  {Feibelman}, \citenamefont {Hammer}, \citenamefont {N{\o}rskov},
  \citenamefont {Wagner}, \citenamefont {Scheffler}, \citenamefont {Stumpf},
  \citenamefont {Watwe},\ and\ \citenamefont {Dumesic}}]{feibelmanCOPt1112001}%
  \BibitemOpen
  \bibfield  {author} {\bibinfo {author} {\bibfnamefont {P.~J.}\ \bibnamefont
  {Feibelman}}, \bibinfo {author} {\bibfnamefont {B.}~\bibnamefont {Hammer}},
  \bibinfo {author} {\bibfnamefont {J.~K.}\ \bibnamefont {N{\o}rskov}},
  \bibinfo {author} {\bibfnamefont {F.}~\bibnamefont {Wagner}}, \bibinfo
  {author} {\bibfnamefont {M.}~\bibnamefont {Scheffler}}, \bibinfo {author}
  {\bibfnamefont {R.}~\bibnamefont {Stumpf}}, \bibinfo {author} {\bibfnamefont
  {R.}~\bibnamefont {Watwe}}, \ and\ \bibinfo {author} {\bibfnamefont
  {J.}~\bibnamefont {Dumesic}},\ }\href {\doibase 10.1021/jp002302t} {\bibfield
   {journal} {\bibinfo  {journal} {The Journal of Physical Chemistry B}\
  }\textbf {\bibinfo {volume} {105}},\ \bibinfo {pages} {4018} (\bibinfo {year}
  {2001})}\BibitemShut {NoStop}%
\bibitem [{\citenamefont {Bao}, \citenamefont {Gagliardi},\ and\ \citenamefont
  {Truhlar}(2018)}]{baoSelfInteractionErrorDensity2018b}%
  \BibitemOpen
  \bibfield  {author} {\bibinfo {author} {\bibfnamefont {J.~L.}\ \bibnamefont
  {Bao}}, \bibinfo {author} {\bibfnamefont {L.}~\bibnamefont {Gagliardi}}, \
  and\ \bibinfo {author} {\bibfnamefont {D.~G.}\ \bibnamefont {Truhlar}},\
  }\href {\doibase 10.1021/acs.jpclett.8b00242} {\bibfield  {journal} {\bibinfo
   {journal} {The Journal of Physical Chemistry Letters}\ }\textbf {\bibinfo
  {volume} {9}},\ \bibinfo {pages} {2353} (\bibinfo {year} {2018})}\BibitemShut
  {NoStop}%
\bibitem [{\citenamefont {Shi}\ \emph {et~al.}(2022)\citenamefont {Shi},
  \citenamefont {Kapil}, \citenamefont {Zen}, \citenamefont {Chen},
  \citenamefont {Alavi},\ and\ \citenamefont {Michaelides}}]{shi_general_2022}%
  \BibitemOpen
  \bibfield  {author} {\bibinfo {author} {\bibfnamefont {B.~X.}\ \bibnamefont
  {Shi}}, \bibinfo {author} {\bibfnamefont {V.}~\bibnamefont {Kapil}}, \bibinfo
  {author} {\bibfnamefont {A.}~\bibnamefont {Zen}}, \bibinfo {author}
  {\bibfnamefont {J.}~\bibnamefont {Chen}}, \bibinfo {author} {\bibfnamefont
  {A.}~\bibnamefont {Alavi}}, \ and\ \bibinfo {author} {\bibfnamefont
  {A.}~\bibnamefont {Michaelides}},\ }\href {\doibase 10.1063/5.0087031}
  {\bibfield  {journal} {\bibinfo  {journal} {The Journal of Chemical Physics}\
  }\textbf {\bibinfo {volume} {156}},\ \bibinfo {pages} {124704} (\bibinfo
  {year} {2022})}\BibitemShut {NoStop}%
\bibitem [{\citenamefont {Bryenton}\ \emph {et~al.}(2023)\citenamefont
  {Bryenton}, \citenamefont {Adeleke}, \citenamefont {Dale},\ and\
  \citenamefont {Johnson}}]{bryentonDelocalizationErrorGreatest2023}%
  \BibitemOpen
  \bibfield  {author} {\bibinfo {author} {\bibfnamefont {K.~R.}\ \bibnamefont
  {Bryenton}}, \bibinfo {author} {\bibfnamefont {A.~A.}\ \bibnamefont
  {Adeleke}}, \bibinfo {author} {\bibfnamefont {S.~G.}\ \bibnamefont {Dale}}, \
  and\ \bibinfo {author} {\bibfnamefont {E.~R.}\ \bibnamefont {Johnson}},\
  }\href {\doibase 10.1002/wcms.1631} {\bibfield  {journal} {\bibinfo
  {journal} {WIREs Computational Molecular Science}\ }\textbf {\bibinfo
  {volume} {13}},\ \bibinfo {pages} {e1631} (\bibinfo {year}
  {2023})}\BibitemShut {NoStop}%
\bibitem [{\citenamefont {Shi}\ \emph {et~al.}(2024)\citenamefont {Shi},
  \citenamefont {Wales}, \citenamefont {Michaelides},\ and\ \citenamefont
  {Myung}}]{shiGoingGoldstandardAttaining2024a}%
  \BibitemOpen
  \bibfield  {author} {\bibinfo {author} {\bibfnamefont {B.~X.}\ \bibnamefont
  {Shi}}, \bibinfo {author} {\bibfnamefont {D.~J.}\ \bibnamefont {Wales}},
  \bibinfo {author} {\bibfnamefont {A.}~\bibnamefont {Michaelides}}, \ and\
  \bibinfo {author} {\bibfnamefont {C.~W.}\ \bibnamefont {Myung}},\ }\href
  {\doibase 10.1021/acs.jctc.4c00379} {\bibfield  {journal} {\bibinfo
  {journal} {Journal of Chemical Theory and Computation}\ }\textbf {\bibinfo
  {volume} {20}},\ \bibinfo {pages} {5306} (\bibinfo {year}
  {2024})}\BibitemShut {NoStop}%
\bibitem [{\citenamefont {Schimka}\ \emph {et~al.}(2010)\citenamefont
  {Schimka}, \citenamefont {Harl}, \citenamefont {Stroppa}, \citenamefont
  {Gr{\"u}neis}, \citenamefont {Marsman}, \citenamefont {Mittendorfer},\ and\
  \citenamefont {Kresse}}]{schimkaAccurateSurfaceAdsorption2010b}%
  \BibitemOpen
  \bibfield  {author} {\bibinfo {author} {\bibfnamefont {L.}~\bibnamefont
  {Schimka}}, \bibinfo {author} {\bibfnamefont {J.}~\bibnamefont {Harl}},
  \bibinfo {author} {\bibfnamefont {A.}~\bibnamefont {Stroppa}}, \bibinfo
  {author} {\bibfnamefont {A.}~\bibnamefont {Gr{\"u}neis}}, \bibinfo {author}
  {\bibfnamefont {M.}~\bibnamefont {Marsman}}, \bibinfo {author} {\bibfnamefont
  {F.}~\bibnamefont {Mittendorfer}}, \ and\ \bibinfo {author} {\bibfnamefont
  {G.}~\bibnamefont {Kresse}},\ }\href {\doibase 10.1038/nmat2806} {\bibfield
  {journal} {\bibinfo  {journal} {Nature Materials}\ }\textbf {\bibinfo
  {volume} {9}},\ \bibinfo {pages} {741} (\bibinfo {year} {2010})}\BibitemShut
  {NoStop}%
\bibitem [{\citenamefont {Yang}\ \emph {et~al.}(2014)\citenamefont {Yang},
  \citenamefont {Hu}, \citenamefont {Usvyat}, \citenamefont {Matthews},
  \citenamefont {Sch{\"u}tz},\ and\ \citenamefont
  {Chan}}]{yangInitioDeterminationCrystalline2014}%
  \BibitemOpen
  \bibfield  {author} {\bibinfo {author} {\bibfnamefont {J.}~\bibnamefont
  {Yang}}, \bibinfo {author} {\bibfnamefont {W.}~\bibnamefont {Hu}}, \bibinfo
  {author} {\bibfnamefont {D.}~\bibnamefont {Usvyat}}, \bibinfo {author}
  {\bibfnamefont {D.}~\bibnamefont {Matthews}}, \bibinfo {author}
  {\bibfnamefont {M.}~\bibnamefont {Sch{\"u}tz}}, \ and\ \bibinfo {author}
  {\bibfnamefont {G.~K.-L.}\ \bibnamefont {Chan}},\ }\href {\doibase
  10.1126/science.1254419} {\bibfield  {journal} {\bibinfo  {journal}
  {Science}\ }\textbf {\bibinfo {volume} {345}},\ \bibinfo {pages} {640}
  (\bibinfo {year} {2014})}\BibitemShut {NoStop}%
\bibitem [{\citenamefont {Sauer}(2019)}]{sauerInitioCalculationsMolecule2019b}%
  \BibitemOpen
  \bibfield  {author} {\bibinfo {author} {\bibfnamefont {J.}~\bibnamefont
  {Sauer}},\ }\href {\doibase 10.1021/acs.accounts.9b00506} {\bibfield
  {journal} {\bibinfo  {journal} {Accounts of Chemical Research}\ }\textbf
  {\bibinfo {volume} {52}},\ \bibinfo {pages} {3502} (\bibinfo {year}
  {2019})}\BibitemShut {NoStop}%
\bibitem [{\citenamefont {Ye}\ and\ \citenamefont
  {C. Berkelbach}(2024)}]{ye_adsorption_2024}%
  \BibitemOpen
  \bibfield  {author} {\bibinfo {author} {\bibfnamefont {H.-Z.}\ \bibnamefont
  {Ye}}\ and\ \bibinfo {author} {\bibfnamefont {T.}~\bibnamefont
  {C. Berkelbach}},\ }\href {\doibase 10.1039/D4FD00041B} {\bibfield
  {journal} {\bibinfo  {journal} {Faraday Discussions}\ }\textbf {\bibinfo
  {volume} {254}},\ \bibinfo {pages} {628} (\bibinfo {year}
  {2024})}\BibitemShut {NoStop}%
\bibitem [{\citenamefont {Della~Pia}\ \emph {et~al.}(2024)\citenamefont
  {Della~Pia}, \citenamefont {Zen}, \citenamefont {Alfè},\ and\ \citenamefont
  {Michaelides}}]{della_pia_how_2024}%
  \BibitemOpen
  \bibfield  {author} {\bibinfo {author} {\bibfnamefont {F.}~\bibnamefont
  {Della~Pia}}, \bibinfo {author} {\bibfnamefont {A.}~\bibnamefont {Zen}},
  \bibinfo {author} {\bibfnamefont {D.}~\bibnamefont {Alfè}}, \ and\ \bibinfo
  {author} {\bibfnamefont {A.}~\bibnamefont {Michaelides}},\ }\href {\doibase
  10.1103/PhysRevLett.133.046401} {\bibfield  {journal} {\bibinfo  {journal}
  {Physical Review Letters}\ }\textbf {\bibinfo {volume} {133}},\ \bibinfo
  {pages} {046401} (\bibinfo {year} {2024})}\BibitemShut {NoStop}%
\bibitem [{\citenamefont {Shi}\ \emph {et~al.}(2025)\citenamefont {Shi},
  \citenamefont {Rosen}, \citenamefont {Schäfer}, \citenamefont {Grüneis},
  \citenamefont {Kapil}, \citenamefont {Zen},\ and\ \citenamefont
  {Michaelides}}]{shi_accurate_2025}%
  \BibitemOpen
  \bibfield  {author} {\bibinfo {author} {\bibfnamefont {B.~X.}\ \bibnamefont
  {Shi}}, \bibinfo {author} {\bibfnamefont {A.~S.}\ \bibnamefont {Rosen}},
  \bibinfo {author} {\bibfnamefont {T.}~\bibnamefont {Schäfer}}, \bibinfo
  {author} {\bibfnamefont {A.}~\bibnamefont {Grüneis}}, \bibinfo {author}
  {\bibfnamefont {V.}~\bibnamefont {Kapil}}, \bibinfo {author} {\bibfnamefont
  {A.}~\bibnamefont {Zen}}, \ and\ \bibinfo {author} {\bibfnamefont
  {A.}~\bibnamefont {Michaelides}},\ }\href {\doibase
  10.1038/s41557-025-01884-y} {\bibfield  {journal} {\bibinfo  {journal}
  {Nature Chemistry}\ ,\ \bibinfo {pages} {1}} (\bibinfo {year}
  {2025})}\BibitemShut {NoStop}%
\bibitem [{\citenamefont {Daru}\ \emph {et~al.}(2022)\citenamefont {Daru},
  \citenamefont {Forbert}, \citenamefont {Behler},\ and\ \citenamefont
  {Marx}}]{daru_coupled_2022}%
  \BibitemOpen
  \bibfield  {author} {\bibinfo {author} {\bibfnamefont {J.}~\bibnamefont
  {Daru}}, \bibinfo {author} {\bibfnamefont {H.}~\bibnamefont {Forbert}},
  \bibinfo {author} {\bibfnamefont {J.}~\bibnamefont {Behler}}, \ and\ \bibinfo
  {author} {\bibfnamefont {D.}~\bibnamefont {Marx}},\ }\href {\doibase
  10.1103/PhysRevLett.129.226001} {\bibfield  {journal} {\bibinfo  {journal}
  {Physical Review Letters}\ }\textbf {\bibinfo {volume} {129}},\ \bibinfo
  {pages} {226001} (\bibinfo {year} {2022})}\BibitemShut {NoStop}%
\bibitem [{\citenamefont {Mészáros}, \citenamefont {Szabó},\ and\
  \citenamefont {Daru}(2025)}]{meszaros_short-range_2025}%
  \BibitemOpen
  \bibfield  {author} {\bibinfo {author} {\bibfnamefont {B.~B.}\ \bibnamefont
  {Mészáros}}, \bibinfo {author} {\bibfnamefont {A.}~\bibnamefont {Szabó}},
  \ and\ \bibinfo {author} {\bibfnamefont {J.}~\bibnamefont {Daru}},\ }\href
  {\doibase 10.1021/acs.jctc.5c00367} {\bibfield  {journal} {\bibinfo
  {journal} {Journal of Chemical Theory and Computation}\ }\textbf {\bibinfo
  {volume} {21}},\ \bibinfo {pages} {5372} (\bibinfo {year}
  {2025})}\BibitemShut {NoStop}%
\bibitem [{\citenamefont {O'Neill}\ \emph {et~al.}(2025)\citenamefont
  {O'Neill}, \citenamefont {Shi}, \citenamefont {Baldwin}, \citenamefont
  {Witt}, \citenamefont {Csányi}, \citenamefont {Gale}, \citenamefont
  {Michaelides},\ and\ \citenamefont {Schran}}]{oneill_towards_2025}%
  \BibitemOpen
  \bibfield  {author} {\bibinfo {author} {\bibfnamefont {N.}~\bibnamefont
  {O'Neill}}, \bibinfo {author} {\bibfnamefont {B.~X.}\ \bibnamefont {Shi}},
  \bibinfo {author} {\bibfnamefont {W.}~\bibnamefont {Baldwin}}, \bibinfo
  {author} {\bibfnamefont {W.~C.}\ \bibnamefont {Witt}}, \bibinfo {author}
  {\bibfnamefont {G.}~\bibnamefont {Csányi}}, \bibinfo {author} {\bibfnamefont
  {J.~D.}\ \bibnamefont {Gale}}, \bibinfo {author} {\bibfnamefont
  {A.}~\bibnamefont {Michaelides}}, \ and\ \bibinfo {author} {\bibfnamefont
  {C.}~\bibnamefont {Schran}},\ }\href {\doibase 10.48550/arXiv.2508.13391}
  {\enquote {\bibinfo {title} {Towards {Routine} {Condensed} {Phase}
  {Simulations} with {Delta}-{Learned} {Coupled} {Cluster} {Accuracy}:
  {Application} to {Liquid} {Water}},}\ } (\bibinfo {year} {2025}),\ \bibinfo
  {note} {arXiv:2508.13391 [physics]}\BibitemShut {NoStop}%
\bibitem [{\citenamefont {Medders}\ \emph {et~al.}(2015)\citenamefont
  {Medders}, \citenamefont {Götz}, \citenamefont {Morales}, \citenamefont
  {Bajaj},\ and\ \citenamefont {Paesani}}]{medders_representation_2015}%
  \BibitemOpen
  \bibfield  {author} {\bibinfo {author} {\bibfnamefont {G.~R.}\ \bibnamefont
  {Medders}}, \bibinfo {author} {\bibfnamefont {A.~W.}\ \bibnamefont {Götz}},
  \bibinfo {author} {\bibfnamefont {M.~A.}\ \bibnamefont {Morales}}, \bibinfo
  {author} {\bibfnamefont {P.}~\bibnamefont {Bajaj}}, \ and\ \bibinfo {author}
  {\bibfnamefont {F.}~\bibnamefont {Paesani}},\ }\href {\doibase
  10.1063/1.4930194} {\bibfield  {journal} {\bibinfo  {journal} {The Journal of
  Chemical Physics}\ }\textbf {\bibinfo {volume} {143}},\ \bibinfo {pages}
  {104102} (\bibinfo {year} {2015})}\BibitemShut {NoStop}%
\bibitem [{\citenamefont {Reddy}\ \emph {et~al.}(2016)\citenamefont {Reddy},
  \citenamefont {Straight}, \citenamefont {Bajaj}, \citenamefont {Huy~Pham},
  \citenamefont {Riera}, \citenamefont {Moberg}, \citenamefont {Morales},
  \citenamefont {Knight}, \citenamefont {G{\"o}tz},\ and\ \citenamefont
  {Paesani}}]{reddyAccuracyMBpolManybody2016}%
  \BibitemOpen
  \bibfield  {author} {\bibinfo {author} {\bibfnamefont {S.~K.}\ \bibnamefont
  {Reddy}}, \bibinfo {author} {\bibfnamefont {S.~C.}\ \bibnamefont {Straight}},
  \bibinfo {author} {\bibfnamefont {P.}~\bibnamefont {Bajaj}}, \bibinfo
  {author} {\bibfnamefont {C.}~\bibnamefont {Huy~Pham}}, \bibinfo {author}
  {\bibfnamefont {M.}~\bibnamefont {Riera}}, \bibinfo {author} {\bibfnamefont
  {D.~R.}\ \bibnamefont {Moberg}}, \bibinfo {author} {\bibfnamefont {M.~A.}\
  \bibnamefont {Morales}}, \bibinfo {author} {\bibfnamefont {C.}~\bibnamefont
  {Knight}}, \bibinfo {author} {\bibfnamefont {A.~W.}\ \bibnamefont
  {G{\"o}tz}}, \ and\ \bibinfo {author} {\bibfnamefont {F.}~\bibnamefont
  {Paesani}},\ }\href {\doibase 10.1063/1.4967719} {\bibfield  {journal}
  {\bibinfo  {journal} {The Journal of Chemical Physics}\ }\textbf {\bibinfo
  {volume} {145}},\ \bibinfo {pages} {194504} (\bibinfo {year}
  {2016})}\BibitemShut {NoStop}%
\bibitem [{\citenamefont {Qu}\ \emph {et~al.}(2023)\citenamefont {Qu},
  \citenamefont {Yu}, \citenamefont {Houston}, \citenamefont {Conte},
  \citenamefont {Nandi},\ and\ \citenamefont
  {Bowman}}]{quInterfacingQAQUAPolarizable2023}%
  \BibitemOpen
  \bibfield  {author} {\bibinfo {author} {\bibfnamefont {C.}~\bibnamefont
  {Qu}}, \bibinfo {author} {\bibfnamefont {Q.}~\bibnamefont {Yu}}, \bibinfo
  {author} {\bibfnamefont {P.~L.}\ \bibnamefont {Houston}}, \bibinfo {author}
  {\bibfnamefont {R.}~\bibnamefont {Conte}}, \bibinfo {author} {\bibfnamefont
  {A.}~\bibnamefont {Nandi}}, \ and\ \bibinfo {author} {\bibfnamefont {J.~M.}\
  \bibnamefont {Bowman}},\ }\href {\doibase 10.1021/acs.jctc.3c00334}
  {\bibfield  {journal} {\bibinfo  {journal} {Journal of Chemical Theory and
  Computation}\ }\textbf {\bibinfo {volume} {19}},\ \bibinfo {pages} {3446}
  (\bibinfo {year} {2023})}\BibitemShut {NoStop}%
\bibitem [{\citenamefont {Xie}\ and\ \citenamefont
  {Bowman}(2010)}]{xie_permutationally_2010}%
  \BibitemOpen
  \bibfield  {author} {\bibinfo {author} {\bibfnamefont {Z.}~\bibnamefont
  {Xie}}\ and\ \bibinfo {author} {\bibfnamefont {J.~M.}\ \bibnamefont
  {Bowman}},\ }\href {\doibase 10.1021/ct9004917} {\bibfield  {journal}
  {\bibinfo  {journal} {Journal of Chemical Theory and Computation}\ }\textbf
  {\bibinfo {volume} {6}},\ \bibinfo {pages} {26} (\bibinfo {year}
  {2010})}\BibitemShut {NoStop}%
\bibitem [{\citenamefont {Lan}\ \emph {et~al.}(2021)\citenamefont {Lan},
  \citenamefont {Kapil}, \citenamefont {Gasparotto}, \citenamefont {Ceriotti},
  \citenamefont {Iannuzzi},\ and\ \citenamefont
  {Rybkin}}]{lan_simulating_2021}%
  \BibitemOpen
  \bibfield  {author} {\bibinfo {author} {\bibfnamefont {J.}~\bibnamefont
  {Lan}}, \bibinfo {author} {\bibfnamefont {V.}~\bibnamefont {Kapil}}, \bibinfo
  {author} {\bibfnamefont {P.}~\bibnamefont {Gasparotto}}, \bibinfo {author}
  {\bibfnamefont {M.}~\bibnamefont {Ceriotti}}, \bibinfo {author}
  {\bibfnamefont {M.}~\bibnamefont {Iannuzzi}}, \ and\ \bibinfo {author}
  {\bibfnamefont {V.~V.}\ \bibnamefont {Rybkin}},\ }\href {\doibase
  10.1038/s41467-021-20914-0} {\bibfield  {journal} {\bibinfo  {journal}
  {Nature Communications}\ }\textbf {\bibinfo {volume} {12}},\ \bibinfo {pages}
  {766} (\bibinfo {year} {2021})}\BibitemShut {NoStop}%
\bibitem [{\citenamefont {Behler}(2015)}]{behler_constructing_2015}%
  \BibitemOpen
  \bibfield  {author} {\bibinfo {author} {\bibfnamefont {J.}~\bibnamefont
  {Behler}},\ }\href {\doibase 10.1002/qua.24890} {\bibfield  {journal}
  {\bibinfo  {journal} {International Journal of Quantum Chemistry}\ }\textbf
  {\bibinfo {volume} {115}},\ \bibinfo {pages} {1032} (\bibinfo {year}
  {2015})}\BibitemShut {NoStop}%
\bibitem [{\citenamefont {Del~Ben}, \citenamefont {Hutter},\ and\ \citenamefont
  {VandeVondele}(2013)}]{del_ben_electron_2013}%
  \BibitemOpen
  \bibfield  {author} {\bibinfo {author} {\bibfnamefont {M.}~\bibnamefont
  {Del~Ben}}, \bibinfo {author} {\bibfnamefont {J.}~\bibnamefont {Hutter}}, \
  and\ \bibinfo {author} {\bibfnamefont {J.}~\bibnamefont {VandeVondele}},\
  }\href {\doibase 10.1021/ct4002202} {\bibfield  {journal} {\bibinfo
  {journal} {Journal of Chemical Theory and Computation}\ }\textbf {\bibinfo
  {volume} {9}},\ \bibinfo {pages} {2654} (\bibinfo {year} {2013})}\BibitemShut
  {NoStop}%
\bibitem [{\citenamefont {Chen}\ \emph {et~al.}(2023)\citenamefont {Chen},
  \citenamefont {Lee}, \citenamefont {Ye}, \citenamefont {Berkelbach},
  \citenamefont {Reichman},\ and\ \citenamefont
  {Markland}}]{chen_data-efficient_2023}%
  \BibitemOpen
  \bibfield  {author} {\bibinfo {author} {\bibfnamefont {M.~S.}\ \bibnamefont
  {Chen}}, \bibinfo {author} {\bibfnamefont {J.}~\bibnamefont {Lee}}, \bibinfo
  {author} {\bibfnamefont {H.-Z.}\ \bibnamefont {Ye}}, \bibinfo {author}
  {\bibfnamefont {T.~C.}\ \bibnamefont {Berkelbach}}, \bibinfo {author}
  {\bibfnamefont {D.~R.}\ \bibnamefont {Reichman}}, \ and\ \bibinfo {author}
  {\bibfnamefont {T.~E.}\ \bibnamefont {Markland}},\ }\href {\doibase
  10.1021/acs.jctc.2c01203} {\bibfield  {journal} {\bibinfo  {journal} {Journal
  of Chemical Theory and Computation}\ }\textbf {\bibinfo {volume} {19}},\
  \bibinfo {pages} {4510} (\bibinfo {year} {2023})}\BibitemShut {NoStop}%
\bibitem [{\citenamefont {Thrun}(1995)}]{Thrun1995}%
  \BibitemOpen
  \bibfield  {author} {\bibinfo {author} {\bibfnamefont {S.}~\bibnamefont
  {Thrun}},\ }in\ \href
  {https://proceedings.neurips.cc/paper_files/paper/1995/file/bdb106a0560c4e46ccc488ef010af787-Paper.pdf}
  {\emph {\bibinfo {booktitle} {Advances in Neural Information Processing
  Systems}}},\ Vol.~\bibinfo {volume} {8},\ \bibinfo {editor} {edited by\
  \bibinfo {editor} {\bibfnamefont {D.~S.}\ \bibnamefont {Touretzky}}, \bibinfo
  {editor} {\bibfnamefont {M.~C.}\ \bibnamefont {Mozer}}, \ and\ \bibinfo
  {editor} {\bibfnamefont {M.~E.}\ \bibnamefont {Hasselmo}}}\ (\bibinfo
  {publisher} {MIT Press},\ \bibinfo {year} {1995})\BibitemShut {NoStop}%
\bibitem [{\citenamefont {Purvis}\ and\ \citenamefont
  {Bartlett}(1982)}]{purvis_full_1982}%
  \BibitemOpen
  \bibfield  {author} {\bibinfo {author} {\bibfnamefont {G.~D.}\ \bibnamefont
  {Purvis}, \bibfnamefont {III}}\ and\ \bibinfo {author} {\bibfnamefont
  {R.~J.}\ \bibnamefont {Bartlett}},\ }\href {\doibase 10.1063/1.443164}
  {\bibfield  {journal} {\bibinfo  {journal} {The Journal of Chemical Physics}\
  }\textbf {\bibinfo {volume} {76}},\ \bibinfo {pages} {1910} (\bibinfo {year}
  {1982})}\BibitemShut {NoStop}%
\bibitem [{\citenamefont {Raghavachari}\ \emph {et~al.}(1989)\citenamefont
  {Raghavachari}, \citenamefont {Trucks}, \citenamefont {Pople},\ and\
  \citenamefont {Head-Gordon}}]{raghavachari_fifth-order_1989}%
  \BibitemOpen
  \bibfield  {author} {\bibinfo {author} {\bibfnamefont {K.}~\bibnamefont
  {Raghavachari}}, \bibinfo {author} {\bibfnamefont {G.~W.}\ \bibnamefont
  {Trucks}}, \bibinfo {author} {\bibfnamefont {J.~A.}\ \bibnamefont {Pople}}, \
  and\ \bibinfo {author} {\bibfnamefont {M.}~\bibnamefont {Head-Gordon}},\
  }\href {\doibase 10.1016/S0009-2614(89)87395-6} {\bibfield  {journal}
  {\bibinfo  {journal} {Chemical Physics Letters}\ }\textbf {\bibinfo {volume}
  {157}},\ \bibinfo {pages} {479} (\bibinfo {year} {1989})}\BibitemShut
  {NoStop}%
\bibitem [{\citenamefont {Zhang}\ and\ \citenamefont
  {Krakauer}(2003)}]{zhang_quantum_2003}%
  \BibitemOpen
  \bibfield  {author} {\bibinfo {author} {\bibfnamefont {S.}~\bibnamefont
  {Zhang}}\ and\ \bibinfo {author} {\bibfnamefont {H.}~\bibnamefont
  {Krakauer}},\ }\href {\doibase 10.1103/PhysRevLett.90.136401} {\bibfield
  {journal} {\bibinfo  {journal} {Physical Review Letters}\ }\textbf {\bibinfo
  {volume} {90}},\ \bibinfo {pages} {136401} (\bibinfo {year}
  {2003})}\BibitemShut {NoStop}%
\bibitem [{\citenamefont {Ramakrishnan}\ \emph {et~al.}(2015)\citenamefont
  {Ramakrishnan}, \citenamefont {Dral}, \citenamefont {Rupp},\ and\
  \citenamefont {von Lilienfeld}}]{ramakrishnan_big_2015}%
  \BibitemOpen
  \bibfield  {author} {\bibinfo {author} {\bibfnamefont {R.}~\bibnamefont
  {Ramakrishnan}}, \bibinfo {author} {\bibfnamefont {P.~O.}\ \bibnamefont
  {Dral}}, \bibinfo {author} {\bibfnamefont {M.}~\bibnamefont {Rupp}}, \ and\
  \bibinfo {author} {\bibfnamefont {O.~A.}\ \bibnamefont {von Lilienfeld}},\
  }\href {\doibase 10.1021/acs.jctc.5b00099} {\bibfield  {journal} {\bibinfo
  {journal} {Journal of Chemical Theory and Computation}\ }\textbf {\bibinfo
  {volume} {11}},\ \bibinfo {pages} {2087} (\bibinfo {year}
  {2015})}\BibitemShut {NoStop}%
\bibitem [{\citenamefont {Pinski}\ \emph {et~al.}(2015)\citenamefont {Pinski},
  \citenamefont {Riplinger}, \citenamefont {Valeev},\ and\ \citenamefont
  {Neese}}]{pinski_sparse_2015}%
  \BibitemOpen
  \bibfield  {author} {\bibinfo {author} {\bibfnamefont {P.}~\bibnamefont
  {Pinski}}, \bibinfo {author} {\bibfnamefont {C.}~\bibnamefont {Riplinger}},
  \bibinfo {author} {\bibfnamefont {E.~F.}\ \bibnamefont {Valeev}}, \ and\
  \bibinfo {author} {\bibfnamefont {F.}~\bibnamefont {Neese}},\ }\href
  {\doibase 10.1063/1.4926879} {\bibfield  {journal} {\bibinfo  {journal} {The
  Journal of Chemical Physics}\ }\textbf {\bibinfo {volume} {143}},\ \bibinfo
  {pages} {034108} (\bibinfo {year} {2015})}\BibitemShut {NoStop}%
\bibitem [{\citenamefont {Riplinger}\ \emph {et~al.}(2016)\citenamefont
  {Riplinger}, \citenamefont {Pinski}, \citenamefont {Becker}, \citenamefont
  {Valeev},\ and\ \citenamefont {Neese}}]{riplinger_sparse_2016}%
  \BibitemOpen
  \bibfield  {author} {\bibinfo {author} {\bibfnamefont {C.}~\bibnamefont
  {Riplinger}}, \bibinfo {author} {\bibfnamefont {P.}~\bibnamefont {Pinski}},
  \bibinfo {author} {\bibfnamefont {U.}~\bibnamefont {Becker}}, \bibinfo
  {author} {\bibfnamefont {E.~F.}\ \bibnamefont {Valeev}}, \ and\ \bibinfo
  {author} {\bibfnamefont {F.}~\bibnamefont {Neese}},\ }\href {\doibase
  10.1063/1.4939030} {\bibfield  {journal} {\bibinfo  {journal} {The Journal of
  Chemical Physics}\ }\textbf {\bibinfo {volume} {144}},\ \bibinfo {pages}
  {024109} (\bibinfo {year} {2016})}\BibitemShut {NoStop}%
\bibitem [{\citenamefont {Nigam}, \citenamefont {Pozdnyakov},\ and\
  \citenamefont {Ceriotti}(2020)}]{nigam_recursive_2020}%
  \BibitemOpen
  \bibfield  {author} {\bibinfo {author} {\bibfnamefont {J.}~\bibnamefont
  {Nigam}}, \bibinfo {author} {\bibfnamefont {S.}~\bibnamefont {Pozdnyakov}}, \
  and\ \bibinfo {author} {\bibfnamefont {M.}~\bibnamefont {Ceriotti}},\ }\href
  {\doibase 10.1063/5.0021116} {\bibfield  {journal} {\bibinfo  {journal} {The
  Journal of Chemical Physics}\ }\textbf {\bibinfo {volume} {153}},\ \bibinfo
  {pages} {121101} (\bibinfo {year} {2020})}\BibitemShut {NoStop}%
\bibitem [{\citenamefont {Drautz}(2019)}]{drautz_atomic_2019}%
  \BibitemOpen
  \bibfield  {author} {\bibinfo {author} {\bibfnamefont {R.}~\bibnamefont
  {Drautz}},\ }\href {\doibase 10.1103/PhysRevB.99.014104} {\bibfield
  {journal} {\bibinfo  {journal} {Physical Review B}\ }\textbf {\bibinfo
  {volume} {99}},\ \bibinfo {pages} {014104} (\bibinfo {year}
  {2019})}\BibitemShut {NoStop}%
\bibitem [{\citenamefont {Batzner}\ \emph {et~al.}(2022)\citenamefont
  {Batzner}, \citenamefont {Musaelian}, \citenamefont {Sun}, \citenamefont
  {Geiger}, \citenamefont {Mailoa}, \citenamefont {Kornbluth}, \citenamefont
  {Molinari}, \citenamefont {Smidt},\ and\ \citenamefont {Kozinsky}}]{nequip}%
  \BibitemOpen
  \bibfield  {author} {\bibinfo {author} {\bibfnamefont {S.}~\bibnamefont
  {Batzner}}, \bibinfo {author} {\bibfnamefont {A.}~\bibnamefont {Musaelian}},
  \bibinfo {author} {\bibfnamefont {L.}~\bibnamefont {Sun}}, \bibinfo {author}
  {\bibfnamefont {M.}~\bibnamefont {Geiger}}, \bibinfo {author} {\bibfnamefont
  {J.~P.}\ \bibnamefont {Mailoa}}, \bibinfo {author} {\bibfnamefont
  {M.}~\bibnamefont {Kornbluth}}, \bibinfo {author} {\bibfnamefont
  {N.}~\bibnamefont {Molinari}}, \bibinfo {author} {\bibfnamefont {T.~E.}\
  \bibnamefont {Smidt}}, \ and\ \bibinfo {author} {\bibfnamefont
  {B.}~\bibnamefont {Kozinsky}},\ }\href {\doibase 10.1038/s41467-022-29939-5}
  {\bibfield  {journal} {\bibinfo  {journal} {Nature Communications}\ }\textbf
  {\bibinfo {volume} {13}},\ \bibinfo {pages} {2453} (\bibinfo {year}
  {2022})}\BibitemShut {NoStop}%
\bibitem [{\citenamefont {Batatia}\ \emph
  {et~al.}(2022{\natexlab{a}})\citenamefont {Batatia}, \citenamefont {Batzner},
  \citenamefont {Kovács}, \citenamefont {Musaelian}, \citenamefont {Simm},
  \citenamefont {Drautz}, \citenamefont {Ortner}, \citenamefont {Kozinsky},\
  and\ \citenamefont {Csányi}}]{batatia_design_2022}%
  \BibitemOpen
  \bibfield  {author} {\bibinfo {author} {\bibfnamefont {I.}~\bibnamefont
  {Batatia}}, \bibinfo {author} {\bibfnamefont {S.}~\bibnamefont {Batzner}},
  \bibinfo {author} {\bibfnamefont {D.~P.}\ \bibnamefont {Kovács}}, \bibinfo
  {author} {\bibfnamefont {A.}~\bibnamefont {Musaelian}}, \bibinfo {author}
  {\bibfnamefont {G.~N.~C.}\ \bibnamefont {Simm}}, \bibinfo {author}
  {\bibfnamefont {R.}~\bibnamefont {Drautz}}, \bibinfo {author} {\bibfnamefont
  {C.}~\bibnamefont {Ortner}}, \bibinfo {author} {\bibfnamefont
  {B.}~\bibnamefont {Kozinsky}}, \ and\ \bibinfo {author} {\bibfnamefont
  {G.}~\bibnamefont {Csányi}},\ }\href {\doibase 10.48550/ARXIV.2205.06643} {\
   (\bibinfo {year} {2022}{\natexlab{a}}),\
  10.48550/ARXIV.2205.06643}\BibitemShut {NoStop}%
\bibitem [{\citenamefont {Bochkarev}, \citenamefont {Lysogorskiy},\ and\
  \citenamefont {Drautz}(2024)}]{bochkarev_graph_2024}%
  \BibitemOpen
  \bibfield  {author} {\bibinfo {author} {\bibfnamefont {A.}~\bibnamefont
  {Bochkarev}}, \bibinfo {author} {\bibfnamefont {Y.}~\bibnamefont
  {Lysogorskiy}}, \ and\ \bibinfo {author} {\bibfnamefont {R.}~\bibnamefont
  {Drautz}},\ }\href {\doibase 10.1103/PhysRevX.14.021036} {\bibfield
  {journal} {\bibinfo  {journal} {Physical Review X}\ }\textbf {\bibinfo
  {volume} {14}},\ \bibinfo {pages} {021036} (\bibinfo {year}
  {2024})}\BibitemShut {NoStop}%
\bibitem [{\citenamefont {Darby}\ \emph {et~al.}(2023)\citenamefont {Darby},
  \citenamefont {Kovács}, \citenamefont {Batatia}, \citenamefont {Caro},
  \citenamefont {Hart}, \citenamefont {Ortner},\ and\ \citenamefont
  {Csányi}}]{darby_tensor-reduced_2023}%
  \BibitemOpen
  \bibfield  {author} {\bibinfo {author} {\bibfnamefont {J.~P.}\ \bibnamefont
  {Darby}}, \bibinfo {author} {\bibfnamefont {D.~P.}\ \bibnamefont {Kovács}},
  \bibinfo {author} {\bibfnamefont {I.}~\bibnamefont {Batatia}}, \bibinfo
  {author} {\bibfnamefont {M.~A.}\ \bibnamefont {Caro}}, \bibinfo {author}
  {\bibfnamefont {G.~L.}\ \bibnamefont {Hart}}, \bibinfo {author}
  {\bibfnamefont {C.}~\bibnamefont {Ortner}}, \ and\ \bibinfo {author}
  {\bibfnamefont {G.}~\bibnamefont {Csányi}},\ }\href {\doibase
  10.1103/PhysRevLett.131.028001} {\bibfield  {journal} {\bibinfo  {journal}
  {Physical Review Letters}\ }\textbf {\bibinfo {volume} {131}},\ \bibinfo
  {pages} {028001} (\bibinfo {year} {2023})}\BibitemShut {NoStop}%
\bibitem [{\citenamefont {Kaur}\ \emph {et~al.}(2024)\citenamefont {Kaur},
  \citenamefont {Pia}, \citenamefont {Batatia}, \citenamefont {Advincula},
  \citenamefont {Shi}, \citenamefont {Lan}, \citenamefont {Csányi},
  \citenamefont {Michaelides},\ and\ \citenamefont
  {Kapil}}]{kaur_data-efficient_2024}%
  \BibitemOpen
  \bibfield  {author} {\bibinfo {author} {\bibfnamefont {H.}~\bibnamefont
  {Kaur}}, \bibinfo {author} {\bibfnamefont {F.~D.}\ \bibnamefont {Pia}},
  \bibinfo {author} {\bibfnamefont {I.}~\bibnamefont {Batatia}}, \bibinfo
  {author} {\bibfnamefont {X.~R.}\ \bibnamefont {Advincula}}, \bibinfo {author}
  {\bibfnamefont {B.~X.}\ \bibnamefont {Shi}}, \bibinfo {author} {\bibfnamefont
  {J.}~\bibnamefont {Lan}}, \bibinfo {author} {\bibfnamefont {G.}~\bibnamefont
  {Csányi}}, \bibinfo {author} {\bibfnamefont {A.}~\bibnamefont
  {Michaelides}}, \ and\ \bibinfo {author} {\bibfnamefont {V.}~\bibnamefont
  {Kapil}},\ }\href {\doibase 10.1039/D4FD00107A} {\bibfield  {journal}
  {\bibinfo  {journal} {Faraday Discussions}\ } (\bibinfo {year} {2024}),\
  10.1039/D4FD00107A}\BibitemShut {NoStop}%
\bibitem [{\citenamefont
  {Deng}(2023)}]{https://doi.org/10.6084/m9.figshare.23713842}%
  \BibitemOpen
  \bibfield  {author} {\bibinfo {author} {\bibfnamefont {B.}~\bibnamefont
  {Deng}},\ }\href {\doibase 10.6084/M9.FIGSHARE.23713842} {\enquote {\bibinfo
  {title} {Materials project trajectory (mptrj) dataset},}\ } (\bibinfo {year}
  {2023})\BibitemShut {NoStop}%
\bibitem [{\citenamefont {Bocus}, \citenamefont {Vandenhaute},\ and\
  \citenamefont {Van~Speybroeck}(2025)}]{bocus_operando_2025}%
  \BibitemOpen
  \bibfield  {author} {\bibinfo {author} {\bibfnamefont {M.}~\bibnamefont
  {Bocus}}, \bibinfo {author} {\bibfnamefont {S.}~\bibnamefont {Vandenhaute}},
  \ and\ \bibinfo {author} {\bibfnamefont {V.}~\bibnamefont {Van~Speybroeck}},\
  }\href {\doibase 10.1002/anie.202413637} {\bibfield  {journal} {\bibinfo
  {journal} {Angewandte Chemie International Edition}\ }\textbf {\bibinfo
  {volume} {64}},\ \bibinfo {pages} {e202413637} (\bibinfo {year}
  {2025})}\BibitemShut {NoStop}%
\bibitem [{\citenamefont {Novelli}\ \emph {et~al.}(2025)\citenamefont
  {Novelli}, \citenamefont {Meanti}, \citenamefont {Buigues}, \citenamefont
  {Rosasco}, \citenamefont {Parrinello}, \citenamefont {Pontil},\ and\
  \citenamefont {Bonati}}]{novelli_fast_2025}%
  \BibitemOpen
  \bibfield  {author} {\bibinfo {author} {\bibfnamefont {P.}~\bibnamefont
  {Novelli}}, \bibinfo {author} {\bibfnamefont {G.}~\bibnamefont {Meanti}},
  \bibinfo {author} {\bibfnamefont {P.~J.}\ \bibnamefont {Buigues}}, \bibinfo
  {author} {\bibfnamefont {L.}~\bibnamefont {Rosasco}}, \bibinfo {author}
  {\bibfnamefont {M.}~\bibnamefont {Parrinello}}, \bibinfo {author}
  {\bibfnamefont {M.}~\bibnamefont {Pontil}}, \ and\ \bibinfo {author}
  {\bibfnamefont {L.}~\bibnamefont {Bonati}},\ }\href {\doibase
  10.48550/arXiv.2505.05652} {\enquote {\bibinfo {title} {Fast and {Fourier}
  {Features} for {Transfer} {Learning} of {Interatomic} {Potentials}},}\ }
  (\bibinfo {year} {2025}),\ \bibinfo {note} {arXiv:2505.05652
  [physics]}\BibitemShut {NoStop}%
\bibitem [{\citenamefont {Radova}\ \emph {et~al.}(2025)\citenamefont {Radova},
  \citenamefont {Stark}, \citenamefont {Allen}, \citenamefont {Maurer},\ and\
  \citenamefont {Bartók}}]{radova_fine-tuning_2025}%
  \BibitemOpen
  \bibfield  {author} {\bibinfo {author} {\bibfnamefont {M.}~\bibnamefont
  {Radova}}, \bibinfo {author} {\bibfnamefont {W.~G.}\ \bibnamefont {Stark}},
  \bibinfo {author} {\bibfnamefont {C.~S.}\ \bibnamefont {Allen}}, \bibinfo
  {author} {\bibfnamefont {R.~J.}\ \bibnamefont {Maurer}}, \ and\ \bibinfo
  {author} {\bibfnamefont {A.~P.}\ \bibnamefont {Bartók}},\ }\href {\doibase
  10.48550/arXiv.2502.15582} {\enquote {\bibinfo {title} {Fine-tuning
  foundation models of materials interatomic potentials with frozen transfer
  learning},}\ } (\bibinfo {year} {2025}),\ \bibinfo {note} {arXiv:2502.15582
  [cond-mat]}\BibitemShut {NoStop}%
\bibitem [{\citenamefont {Mazitov}\ \emph {et~al.}(2025)\citenamefont
  {Mazitov}, \citenamefont {Bigi}, \citenamefont {Kellner}, \citenamefont
  {Pegolo}, \citenamefont {Tisi}, \citenamefont {Fraux}, \citenamefont
  {Pozdnyakov}, \citenamefont {Loche},\ and\ \citenamefont
  {Ceriotti}}]{mazitov_pet-mad_2025}%
  \BibitemOpen
  \bibfield  {author} {\bibinfo {author} {\bibfnamefont {A.}~\bibnamefont
  {Mazitov}}, \bibinfo {author} {\bibfnamefont {F.}~\bibnamefont {Bigi}},
  \bibinfo {author} {\bibfnamefont {M.}~\bibnamefont {Kellner}}, \bibinfo
  {author} {\bibfnamefont {P.}~\bibnamefont {Pegolo}}, \bibinfo {author}
  {\bibfnamefont {D.}~\bibnamefont {Tisi}}, \bibinfo {author} {\bibfnamefont
  {G.}~\bibnamefont {Fraux}}, \bibinfo {author} {\bibfnamefont
  {S.}~\bibnamefont {Pozdnyakov}}, \bibinfo {author} {\bibfnamefont
  {P.}~\bibnamefont {Loche}}, \ and\ \bibinfo {author} {\bibfnamefont
  {M.}~\bibnamefont {Ceriotti}},\ }\href {\doibase 10.48550/arXiv.2503.14118}
  {\enquote {\bibinfo {title} {{PET}-{MAD}, a lightweight universal interatomic
  potential for advanced materials modeling},}\ } (\bibinfo {year} {2025}),\
  \bibinfo {note} {arXiv:2503.14118 [cond-mat]}\BibitemShut {NoStop}%
\bibitem [{\citenamefont {Messerly}\ \emph {et~al.}(2025)\citenamefont
  {Messerly}, \citenamefont {Matin}, \citenamefont {Allen}, \citenamefont
  {Nebgen}, \citenamefont {Barros}, \citenamefont {Smith}, \citenamefont
  {Lubbers},\ and\ \citenamefont {Messerly}}]{messerly_multi-fidelity_2025}%
  \BibitemOpen
  \bibfield  {author} {\bibinfo {author} {\bibfnamefont {M.}~\bibnamefont
  {Messerly}}, \bibinfo {author} {\bibfnamefont {S.}~\bibnamefont {Matin}},
  \bibinfo {author} {\bibfnamefont {A.~E.~A.}\ \bibnamefont {Allen}}, \bibinfo
  {author} {\bibfnamefont {B.}~\bibnamefont {Nebgen}}, \bibinfo {author}
  {\bibfnamefont {K.}~\bibnamefont {Barros}}, \bibinfo {author} {\bibfnamefont
  {J.~S.}\ \bibnamefont {Smith}}, \bibinfo {author} {\bibfnamefont
  {N.}~\bibnamefont {Lubbers}}, \ and\ \bibinfo {author} {\bibfnamefont
  {R.}~\bibnamefont {Messerly}},\ }\href {\doibase 10.48550/arXiv.2505.01590}
  {\enquote {\bibinfo {title} {Multi-fidelity learning for interatomic
  potentials: {Low}-level forces and high-level energies are all you need},}\ }
  (\bibinfo {year} {2025}),\ \bibinfo {note} {arXiv:2505.01590
  [physics]}\BibitemShut {NoStop}%
\bibitem [{\citenamefont {Cui}, \citenamefont {Reuter},\ and\ \citenamefont
  {Margraf}(2025)}]{cui_multi-fidelity_2025}%
  \BibitemOpen
  \bibfield  {author} {\bibinfo {author} {\bibfnamefont {M.}~\bibnamefont
  {Cui}}, \bibinfo {author} {\bibfnamefont {K.}~\bibnamefont {Reuter}}, \ and\
  \bibinfo {author} {\bibfnamefont {J.~T.}\ \bibnamefont {Margraf}},\ }\href
  {\doibase 10.1088/2632-2153/adc222} {\bibfield  {journal} {\bibinfo
  {journal} {Machine Learning: Science and Technology}\ }\textbf {\bibinfo
  {volume} {6}},\ \bibinfo {pages} {015071} (\bibinfo {year}
  {2025})}\BibitemShut {NoStop}%
\bibitem [{\citenamefont {Pia}\ \emph {et~al.}(2025)\citenamefont {Pia},
  \citenamefont {Shi}, \citenamefont {Kapil}, \citenamefont {Zen},
  \citenamefont {Alfè},\ and\ \citenamefont
  {Michaelides}}]{pia_accurate_2025}%
  \BibitemOpen
  \bibfield  {author} {\bibinfo {author} {\bibfnamefont {F.~D.}\ \bibnamefont
  {Pia}}, \bibinfo {author} {\bibfnamefont {B.~X.}\ \bibnamefont {Shi}},
  \bibinfo {author} {\bibfnamefont {V.}~\bibnamefont {Kapil}}, \bibinfo
  {author} {\bibfnamefont {A.}~\bibnamefont {Zen}}, \bibinfo {author}
  {\bibfnamefont {D.}~\bibnamefont {Alfè}}, \ and\ \bibinfo {author}
  {\bibfnamefont {A.}~\bibnamefont {Michaelides}},\ }\href {\doibase
  10.1039/D5SC01325A} {\bibfield  {journal} {\bibinfo  {journal} {Chemical
  Science}\ }\textbf {\bibinfo {volume} {16}},\ \bibinfo {pages} {11419}
  (\bibinfo {year} {2025})}\BibitemShut {NoStop}%
\bibitem [{\citenamefont {Tuckerman}(2010)}]{tuckerman_statistical_2010}%
  \BibitemOpen
  \bibfield  {author} {\bibinfo {author} {\bibfnamefont {M.}~\bibnamefont
  {Tuckerman}},\ }\href@noop {} {{\selectlanguage {en}\emph {\bibinfo {title}
  {Statistical {Mechanics}: {Theory} and {Molecular} {Simulation}}}}}\
  (\bibinfo  {publisher} {OUP Oxford},\ \bibinfo {year} {2010})\ \bibinfo
  {note} {google-Books-ID: Lo3Jqc0pgrcC}\BibitemShut {NoStop}%
\bibitem [{\citenamefont {Batatia}\ \emph
  {et~al.}(2022{\natexlab{b}})\citenamefont {Batatia}, \citenamefont {Kovacs},
  \citenamefont {Simm}, \citenamefont {Ortner},\ and\ \citenamefont
  {Csanyi}}]{batatia_mace_2022}%
  \BibitemOpen
  \bibfield  {author} {\bibinfo {author} {\bibfnamefont {I.}~\bibnamefont
  {Batatia}}, \bibinfo {author} {\bibfnamefont {D.~P.}\ \bibnamefont {Kovacs}},
  \bibinfo {author} {\bibfnamefont {G.}~\bibnamefont {Simm}}, \bibinfo {author}
  {\bibfnamefont {C.}~\bibnamefont {Ortner}}, \ and\ \bibinfo {author}
  {\bibfnamefont {G.}~\bibnamefont {Csanyi}},\ }\href
  {https://proceedings.neurips.cc/paper\_files/paper/2022/hash/4a36c3c51af11ed9f34615b81edb5bbc-Abstract-Conference.html}
  {\bibfield  {journal} {\bibinfo  {journal} {Advances in Neural Information
  Processing Systems}\ }\textbf {\bibinfo {volume} {35}},\ \bibinfo {pages}
  {11423} (\bibinfo {year} {2022}{\natexlab{b}})}\BibitemShut {NoStop}%
\bibitem [{\citenamefont {Litman}\ \emph {et~al.}(2024)\citenamefont {Litman},
  \citenamefont {Kapil}, \citenamefont {Feldman}, \citenamefont {Tisi},
  \citenamefont {Begušić}, \citenamefont {Fidanyan}, \citenamefont {Fraux},
  \citenamefont {Higer}, \citenamefont {Kellner}, \citenamefont {Li},
  \citenamefont {Pós}, \citenamefont {Stocco}, \citenamefont {Trenins},
  \citenamefont {Hirshberg}, \citenamefont {Rossi},\ and\ \citenamefont
  {Ceriotti}}]{ipi}%
  \BibitemOpen
  \bibfield  {author} {\bibinfo {author} {\bibfnamefont {Y.}~\bibnamefont
  {Litman}}, \bibinfo {author} {\bibfnamefont {V.}~\bibnamefont {Kapil}},
  \bibinfo {author} {\bibfnamefont {Y.~M.~Y.}\ \bibnamefont {Feldman}},
  \bibinfo {author} {\bibfnamefont {D.}~\bibnamefont {Tisi}}, \bibinfo {author}
  {\bibfnamefont {T.}~\bibnamefont {Begušić}}, \bibinfo {author}
  {\bibfnamefont {K.}~\bibnamefont {Fidanyan}}, \bibinfo {author}
  {\bibfnamefont {G.}~\bibnamefont {Fraux}}, \bibinfo {author} {\bibfnamefont
  {J.}~\bibnamefont {Higer}}, \bibinfo {author} {\bibfnamefont
  {M.}~\bibnamefont {Kellner}}, \bibinfo {author} {\bibfnamefont {T.~E.}\
  \bibnamefont {Li}}, \bibinfo {author} {\bibfnamefont {E.~S.}\ \bibnamefont
  {Pós}}, \bibinfo {author} {\bibfnamefont {E.}~\bibnamefont {Stocco}},
  \bibinfo {author} {\bibfnamefont {G.}~\bibnamefont {Trenins}}, \bibinfo
  {author} {\bibfnamefont {B.}~\bibnamefont {Hirshberg}}, \bibinfo {author}
  {\bibfnamefont {M.}~\bibnamefont {Rossi}}, \ and\ \bibinfo {author}
  {\bibfnamefont {M.}~\bibnamefont {Ceriotti}},\ }\href {\doibase
  10.1063/5.0215869} {\bibfield  {journal} {\bibinfo  {journal} {The Journal of
  Chemical Physics}\ }\textbf {\bibinfo {volume} {161}},\ \bibinfo {pages}
  {062504} (\bibinfo {year} {2024})},\ \Eprint
  {http://arxiv.org/abs/https://pubs.aip.org/aip/jcp/article-pdf/doi/10.1063/5.0215869/20111898/062504\_1\_5.0215869.pdf}
  {https://pubs.aip.org/aip/jcp/article-pdf/doi/10.1063/5.0215869/20111898/062504\_1\_5.0215869.pdf}
  \BibitemShut {NoStop}%
\bibitem [{\citenamefont {Hjorth~Larsen}\ \emph {et~al.}(2017)\citenamefont
  {Hjorth~Larsen}, \citenamefont {J{\o}rgen~Mortensen}, \citenamefont
  {Blomqvist}, \citenamefont {Castelli}, \citenamefont {Christensen},
  \citenamefont {Du{\l}ak}, \citenamefont {Friis}, \citenamefont {Groves},
  \citenamefont {Hammer}, \citenamefont {Hargus}, \citenamefont {Hermes},
  \citenamefont {Jennings}, \citenamefont {Bjerre~Jensen}, \citenamefont
  {Kermode}, \citenamefont {Kitchin}, \citenamefont {Leonhard~Kolsbjerg},
  \citenamefont {Kubal}, \citenamefont {Kaasbjerg}, \citenamefont {Lysgaard},
  \citenamefont {Bergmann~Maronsson}, \citenamefont {Maxson}, \citenamefont
  {Olsen}, \citenamefont {Pastewka}, \citenamefont {Peterson}, \citenamefont
  {Rostgaard}, \citenamefont {Schi{\o}tz}, \citenamefont {Sch{\"u}tt},
  \citenamefont {Strange}, \citenamefont {Thygesen}, \citenamefont {Vegge},
  \citenamefont {Vilhelmsen}, \citenamefont {Walter}, \citenamefont {Zeng},\
  and\ \citenamefont {Jacobsen}}]{ase}%
  \BibitemOpen
  \bibfield  {author} {\bibinfo {author} {\bibfnamefont {A.}~\bibnamefont
  {Hjorth~Larsen}}, \bibinfo {author} {\bibfnamefont {J.}~\bibnamefont
  {J{\o}rgen~Mortensen}}, \bibinfo {author} {\bibfnamefont {J.}~\bibnamefont
  {Blomqvist}}, \bibinfo {author} {\bibfnamefont {I.~E.}\ \bibnamefont
  {Castelli}}, \bibinfo {author} {\bibfnamefont {R.}~\bibnamefont
  {Christensen}}, \bibinfo {author} {\bibfnamefont {M.}~\bibnamefont
  {Du{\l}ak}}, \bibinfo {author} {\bibfnamefont {J.}~\bibnamefont {Friis}},
  \bibinfo {author} {\bibfnamefont {M.~N.}\ \bibnamefont {Groves}}, \bibinfo
  {author} {\bibfnamefont {B.}~\bibnamefont {Hammer}}, \bibinfo {author}
  {\bibfnamefont {C.}~\bibnamefont {Hargus}}, \bibinfo {author} {\bibfnamefont
  {E.~D.}\ \bibnamefont {Hermes}}, \bibinfo {author} {\bibfnamefont {P.~C.}\
  \bibnamefont {Jennings}}, \bibinfo {author} {\bibfnamefont {P.}~\bibnamefont
  {Bjerre~Jensen}}, \bibinfo {author} {\bibfnamefont {J.}~\bibnamefont
  {Kermode}}, \bibinfo {author} {\bibfnamefont {J.~R.}\ \bibnamefont
  {Kitchin}}, \bibinfo {author} {\bibfnamefont {E.}~\bibnamefont
  {Leonhard~Kolsbjerg}}, \bibinfo {author} {\bibfnamefont {J.}~\bibnamefont
  {Kubal}}, \bibinfo {author} {\bibfnamefont {K.}~\bibnamefont {Kaasbjerg}},
  \bibinfo {author} {\bibfnamefont {S.}~\bibnamefont {Lysgaard}}, \bibinfo
  {author} {\bibfnamefont {J.}~\bibnamefont {Bergmann~Maronsson}}, \bibinfo
  {author} {\bibfnamefont {T.}~\bibnamefont {Maxson}}, \bibinfo {author}
  {\bibfnamefont {T.}~\bibnamefont {Olsen}}, \bibinfo {author} {\bibfnamefont
  {L.}~\bibnamefont {Pastewka}}, \bibinfo {author} {\bibfnamefont
  {A.}~\bibnamefont {Peterson}}, \bibinfo {author} {\bibfnamefont
  {C.}~\bibnamefont {Rostgaard}}, \bibinfo {author} {\bibfnamefont
  {J.}~\bibnamefont {Schi{\o}tz}}, \bibinfo {author} {\bibfnamefont
  {O.}~\bibnamefont {Sch{\"u}tt}}, \bibinfo {author} {\bibfnamefont
  {M.}~\bibnamefont {Strange}}, \bibinfo {author} {\bibfnamefont {K.~S.}\
  \bibnamefont {Thygesen}}, \bibinfo {author} {\bibfnamefont {T.}~\bibnamefont
  {Vegge}}, \bibinfo {author} {\bibfnamefont {L.}~\bibnamefont {Vilhelmsen}},
  \bibinfo {author} {\bibfnamefont {M.}~\bibnamefont {Walter}}, \bibinfo
  {author} {\bibfnamefont {Z.}~\bibnamefont {Zeng}}, \ and\ \bibinfo {author}
  {\bibfnamefont {K.~W.}\ \bibnamefont {Jacobsen}},\ }\href@noop {} {\bibfield
  {journal} {\bibinfo  {journal} {J. Phys. Condens. Matter}\ }\textbf {\bibinfo
  {volume} {29}},\ \bibinfo {pages} {273002} (\bibinfo {year}
  {2017})}\BibitemShut {NoStop}%
\bibitem [{\citenamefont {Stolte}\ \emph {et~al.}(2024)\citenamefont {Stolte},
  \citenamefont {Daru}, \citenamefont {Forbert}, \citenamefont {Behler},\ and\
  \citenamefont {Marx}}]{stolte_nuclear_2024}%
  \BibitemOpen
  \bibfield  {author} {\bibinfo {author} {\bibfnamefont {N.}~\bibnamefont
  {Stolte}}, \bibinfo {author} {\bibfnamefont {J.}~\bibnamefont {Daru}},
  \bibinfo {author} {\bibfnamefont {H.}~\bibnamefont {Forbert}}, \bibinfo
  {author} {\bibfnamefont {J.}~\bibnamefont {Behler}}, \ and\ \bibinfo {author}
  {\bibfnamefont {D.}~\bibnamefont {Marx}},\ }\href {\doibase
  10.1021/acs.jpclett.4c02925} {\bibfield  {journal} {\bibinfo  {journal} {The
  Journal of Physical Chemistry Letters}\ }\textbf {\bibinfo {volume} {15}},\
  \bibinfo {pages} {12144} (\bibinfo {year} {2024})}\BibitemShut {NoStop}%
\bibitem [{\citenamefont {Ceriotti}\ \emph {et~al.}(2013)\citenamefont
  {Ceriotti}, \citenamefont {Cuny}, \citenamefont {Parrinello},\ and\
  \citenamefont {Manolopoulos}}]{ceriotti_nuclear_2013}%
  \BibitemOpen
  \bibfield  {author} {\bibinfo {author} {\bibfnamefont {M.}~\bibnamefont
  {Ceriotti}}, \bibinfo {author} {\bibfnamefont {J.}~\bibnamefont {Cuny}},
  \bibinfo {author} {\bibfnamefont {M.}~\bibnamefont {Parrinello}}, \ and\
  \bibinfo {author} {\bibfnamefont {D.~E.}\ \bibnamefont {Manolopoulos}},\
  }\href {\doibase 10.1073/pnas.1308560110} {\bibfield  {journal} {\bibinfo
  {journal} {Proceedings of the National Academy of Sciences}\ }\textbf
  {\bibinfo {volume} {110}},\ \bibinfo {pages} {15591} (\bibinfo {year}
  {2013})}\BibitemShut {NoStop}%
\bibitem [{\citenamefont {Dasgupta}, \citenamefont {Cassone},\ and\
  \citenamefont {Paesani}(2025)}]{dasgupta_nuclear_2025}%
  \BibitemOpen
  \bibfield  {author} {\bibinfo {author} {\bibfnamefont {S.}~\bibnamefont
  {Dasgupta}}, \bibinfo {author} {\bibfnamefont {G.}~\bibnamefont {Cassone}}, \
  and\ \bibinfo {author} {\bibfnamefont {F.}~\bibnamefont {Paesani}},\ }\href
  {\doibase 10.1021/acs.jpclett.5c00168} {\bibfield  {journal} {\bibinfo
  {journal} {The Journal of Physical Chemistry Letters}\ }\textbf {\bibinfo
  {volume} {16}},\ \bibinfo {pages} {2996} (\bibinfo {year}
  {2025})}\BibitemShut {NoStop}%
\bibitem [{\citenamefont {Kapil}\ \emph {et~al.}(2023)\citenamefont {Kapil},
  \citenamefont {Kovacs}, \citenamefont {Csányi},\ and\ \citenamefont
  {Michaelides}}]{kapil_first-principles_2023}%
  \BibitemOpen
  \bibfield  {author} {\bibinfo {author} {\bibfnamefont {V.}~\bibnamefont
  {Kapil}}, \bibinfo {author} {\bibfnamefont {D.~P.}\ \bibnamefont {Kovacs}},
  \bibinfo {author} {\bibfnamefont {G.}~\bibnamefont {Csányi}}, \ and\
  \bibinfo {author} {\bibfnamefont {A.}~\bibnamefont {Michaelides}},\ }\href
  {\doibase 10.1039/D3FD00113J} {\bibfield  {journal} {\bibinfo  {journal}
  {Faraday Discussions}\ } (\bibinfo {year} {2023}),\
  10.1039/D3FD00113J}\BibitemShut {NoStop}%
\bibitem [{\citenamefont {Bigi}, \citenamefont {Langer},\ and\ \citenamefont
  {Ceriotti}(2025)}]{bigi_dark_2025}%
  \BibitemOpen
  \bibfield  {author} {\bibinfo {author} {\bibfnamefont {F.}~\bibnamefont
  {Bigi}}, \bibinfo {author} {\bibfnamefont {M.}~\bibnamefont {Langer}}, \ and\
  \bibinfo {author} {\bibfnamefont {M.}~\bibnamefont {Ceriotti}},\ }\href
  {\doibase 10.48550/arXiv.2412.11569} {\enquote {\bibinfo {title} {The dark
  side of the forces: assessing non-conservative force models for atomistic
  machine learning},}\ } (\bibinfo {year} {2025}),\ \bibinfo {note}
  {arXiv:2412.11569 [physics]}\BibitemShut {NoStop}%
\bibitem [{\citenamefont {Chong}\ \emph {et~al.}(2025)\citenamefont {Chong},
  \citenamefont {Jiang}, \citenamefont {Domina}, \citenamefont {Bigi},
  \citenamefont {Grasselli}, \citenamefont {Lee},\ and\ \citenamefont
  {Ceriotti}}]{chong_resolving_2025}%
  \BibitemOpen
  \bibfield  {author} {\bibinfo {author} {\bibfnamefont {S.}~\bibnamefont
  {Chong}}, \bibinfo {author} {\bibfnamefont {T.}~\bibnamefont {Jiang}},
  \bibinfo {author} {\bibfnamefont {M.}~\bibnamefont {Domina}}, \bibinfo
  {author} {\bibfnamefont {F.}~\bibnamefont {Bigi}}, \bibinfo {author}
  {\bibfnamefont {F.}~\bibnamefont {Grasselli}}, \bibinfo {author}
  {\bibfnamefont {J.}~\bibnamefont {Lee}}, \ and\ \bibinfo {author}
  {\bibfnamefont {M.}~\bibnamefont {Ceriotti}},\ }\href {\doibase
  10.48550/arXiv.2509.14146} {\enquote {\bibinfo {title} {Resolving the
  {Body}-{Order} {Paradox} of {Machine} {Learning} {Interatomic}
  {Potentials}},}\ } (\bibinfo {year} {2025}),\ \bibinfo {note}
  {arXiv:2509.14146 [physics]}\BibitemShut {NoStop}%
\bibitem [{\citenamefont {Pinski}\ and\ \citenamefont
  {Neese}(2019)}]{pinskiAnalyticalGradientDomainbased2019}%
  \BibitemOpen
  \bibfield  {author} {\bibinfo {author} {\bibfnamefont {P.}~\bibnamefont
  {Pinski}}\ and\ \bibinfo {author} {\bibfnamefont {F.}~\bibnamefont {Neese}},\
  }\href {\doibase 10.1063/1.5086544} {\bibfield  {journal} {\bibinfo
  {journal} {The Journal of Chemical Physics}\ }\textbf {\bibinfo {volume}
  {150}},\ \bibinfo {pages} {164102} (\bibinfo {year} {2019})}\BibitemShut
  {NoStop}%
\bibitem [{\citenamefont {Zhang}\ \emph {et~al.}(2024)\citenamefont {Zhang},
  \citenamefont {Li}, \citenamefont {Ye}, \citenamefont {Berkelbach},\ and\
  \citenamefont {Chan}}]{zhangPerformantAutomaticDifferentiation2024}%
  \BibitemOpen
  \bibfield  {author} {\bibinfo {author} {\bibfnamefont {X.}~\bibnamefont
  {Zhang}}, \bibinfo {author} {\bibfnamefont {C.}~\bibnamefont {Li}}, \bibinfo
  {author} {\bibfnamefont {H.-Z.}\ \bibnamefont {Ye}}, \bibinfo {author}
  {\bibfnamefont {T.~C.}\ \bibnamefont {Berkelbach}}, \ and\ \bibinfo {author}
  {\bibfnamefont {G.~K.-L.}\ \bibnamefont {Chan}},\ }\href {\doibase
  10.48550/arXiv.2404.03129} {\  (\bibinfo {year} {2024}),\
  10.48550/arXiv.2404.03129},\ \Eprint {http://arxiv.org/abs/2404.03129}
  {arXiv:2404.03129 [physics]} \BibitemShut {NoStop}%
\bibitem [{\citenamefont {Slootman}\ \emph {et~al.}(2024)\citenamefont
  {Slootman}, \citenamefont {Poltavsky}, \citenamefont {Shinde}, \citenamefont
  {Cocomello}, \citenamefont {Moroni}, \citenamefont {Tkatchenko},\ and\
  \citenamefont {Filippi}}]{slootmanAccurateQuantumMonte2024}%
  \BibitemOpen
  \bibfield  {author} {\bibinfo {author} {\bibfnamefont {E.}~\bibnamefont
  {Slootman}}, \bibinfo {author} {\bibfnamefont {I.}~\bibnamefont {Poltavsky}},
  \bibinfo {author} {\bibfnamefont {R.}~\bibnamefont {Shinde}}, \bibinfo
  {author} {\bibfnamefont {J.}~\bibnamefont {Cocomello}}, \bibinfo {author}
  {\bibfnamefont {S.}~\bibnamefont {Moroni}}, \bibinfo {author} {\bibfnamefont
  {A.}~\bibnamefont {Tkatchenko}}, \ and\ \bibinfo {author} {\bibfnamefont
  {C.}~\bibnamefont {Filippi}},\ }\href {\doibase 10.1021/acs.jctc.4c00498}
  {\bibfield  {journal} {\bibinfo  {journal} {Journal of Chemical Theory and
  Computation}\ }\textbf {\bibinfo {volume} {20}},\ \bibinfo {pages} {6020}
  (\bibinfo {year} {2024})}\BibitemShut {NoStop}%
\end{thebibliography}
\end{document}